\documentclass[11pt, a4paper]{article}
\pdfoutput=1

\usepackage{amsmath,color,multicol}
\usepackage{amsfonts}
\usepackage{amssymb}
\usepackage{graphicx}
\usepackage{mathrsfs}
\usepackage{epsfig}
\usepackage{latexsym}
\usepackage{graphicx}
\usepackage{color}
\usepackage{cite}
\usepackage{slashed, cancel}
\usepackage{hyperref}
\usepackage{datetime}
\usepackage{cancel}
\usepackage{multirow}

\def\circa#1{\,\raise.3ex\hbox{$#1$\kern-.75em\lower1ex\hbox{$\sim$}}\,}

\newcommand{\beq}{\begin{equation}}
\newcommand{\eeq}{\end{equation}}
\newcommand{\beqn}{\begin{eqnarray}}
\newcommand{\eeqn}{\end{eqnarray}}
\def\app#1#2{%
  \mathrel{%
    \setbox0=\hbox{$#1\sim$}%
    \setbox2=\hbox{%
      \rlap{\hbox{$#1\propto$}}%
      \lower1.1\ht0\box0%
    }%
    \raise0.25\ht2\box2%
  }%
}

\numberwithin{equation}{section}

\font\tenrsfs=rsfs10 at 12pt

\font\sevenrsfs=rsfs7
\font\fiversfs=rsfs5
\newfam\rsfsfam
\textfont\rsfsfam=\tenrsfs
\scriptfont\rsfsfam=\sevenrsfs
\scriptscriptfont\rsfsfam=\fiversfs
\def\mathscr#1{{\fam\rsfsfam\relax#1}}

\hypersetup{colorlinks, citecolor=bluscuro, linkcolor=black, urlcolor=bluscuro}
\definecolor{rossos}{rgb}{0.8,0.2,0.3}
\definecolor{bluscuro}{rgb}{0.15, 0.2, .85}
\definecolor{bluchiaro}{cmyk}{1,.3,0.,0.1}

\makeatother   % Cancel the effect of \makeatletter

 \def\be   {\begin{equation}}   \def\ee   {\end{equation}}
 \def\ba   {\begin{array}}      \def\ea   {\end{array}}
 \def\bea  {\begin{eqnarray}}   \def\eea  {\end{eqnarray}}
 \def\bean {\begin{eqnarray*}}  \def\eean {\end{eqnarray*}}

\begin{document}

%\hfill \texttt{\jobname .tex}\qquad\textit{\currenttime\qquad \dmyyyydate\today}
%
%\vspace{-1cm}

\begin{flushright} 
Preprint CERN-TH-2017-269
\end{flushright}

\newcommand{\stau}{\widetilde{\tau}}
\newcommand{\bino}{\widetilde{B}}
\def\MS{\color{red}} 
~~
\vspace{0.0cm}

\begin{center}

{\LARGE \textbf {Same-sign WW scattering at the LHC:
\\[0.3cm]can we discover BSM effects before discovering new states?
%\\[0.5cm]
% 
}}
\\ [1cm]
%\\ [1.5cm]
{\large {Jan Kalinowski,$^{a,b} $
Pawe{\l} Koz{\'o}w,$^a$ Stefan Pokorski,$^a$
Janusz Rosiek,$^a$ Micha{\l} Szleper$^c$ \\
and
S{\l}awomir Tkaczyk $^d$
}}
\\[1cm]
\end{center}

\hspace{1cm}
\begin{minipage}{14cm}
\textit{ 
\noindent $^a$ Institute of Theoretical Physics, Faculty of Physics, University of Warsaw, 
\\ \phantom{m} ul.~Pasteura 5, PL--02--093 Warsaw, Poland\\[0.1cm]
$^b$ CERN, Theoretical Physics Department,
\\ \phantom{m} CH-1211 Geneva 23, Switzerland\\[0.1cm]
$^c$ National Center for Nuclear Research, High Energy Physics Department,\\
\phantom{m} ul.~Ho\.za 69, PL-00-681, Warsaw, Poland \\[0.1cm]
$^d$  Fermi National Accelarator Laboratory, \\
\phantom{m} Batavia, IL 60510, USA
}
\end{minipage}

%\vspace{-0.3cm}
\vspace{0.7cm}

\begin{center}
\textbf{Abstract}

\begin{quote}
It is possible that measurements of vector boson scattering (VBS) processes
at the LHC will reveal disagreement with Standard Model predictions, but no new particles
will be observed directly.  The task is then to learn as much as possible
about the new physics from a VBS analysis carried within the framework of the
Effective Field Theory (EFT).
In this paper we discuss issues related to the correct usage of the EFT
when the $WW$ invariant mass is not directly accessible experimentally,
as in purely leptonic  $W$ decay channels.
Strategies for future data analyses in case such scenario indeed occurs are proposed.

\end{quote}
%\today
\end{center}
\vspace{10mm}

\def\thefootnote{\arabic{footnote}}
\setcounter{footnote}{0}
\pagestyle{empty}

%\newpage
\pagestyle{plain}
%\setcounter{page}{1}

%%%%%%%%%%%%%%%%%%%%%%%%%%%%%%%%%%%%%%%%%%
\section{Introduction and strategy}
Searches for deviations from Standard Model (SM) predictions in processes involving 
interactions between known particles are a well established technique to study possible
contributions from Beyond the Standard Model (BSM) physics.
In this paper we address the question how much we can learn about the scale of new 
physics and its strength using the Effective Field Theory (EFT) approach 
to $W^+W^+$ scattering if a statistically significant deviation from the SM predictions is observed
in the expected LHC data for the process 
$pp \rightarrow 2 jets + W^+W^+ $.  Our specific focus is on the proper use of the EFT in its 
 range of validity.  With this in mind, we discuss the practical usefulness of
the EFT language to describe vector boson scattering (VBS) data and whether or not this 
can indeed be the right framework to observe the first hints of new physics at the LHC.

The EFT is in principle a model independent tool to describe BSM physics below the 
thresholds for new states.
One supplements the SM Lagrangian by higher dimension operators

\begin{equation}
{\cal L}={\cal L}_{SM} +\Sigma_i\frac{C_i^{(6)}}{\Lambda^2_i}{\cal O}_i^{(6)} +\Sigma_i\frac{C_i^{(8)}}{\Lambda^4_i}{\cal O}_i^{(8)} +...
\label{lagrangian}
\end{equation} 
where the $C$'s are some ``coupling constants" and $\Lambda$'s are the decoupled new mass 
scales.  The mass scale is a feature of the UV completion of the full theory and
thus is assumed common to all the coefficients
\begin{equation}
f_i^{(6)}=\frac{C_i^{(6)}}{\Lambda^2} ,  ~~~~f_i^{(8)}=\frac{C_i^{(8)}}{\Lambda^4},....
\label{efy}
\end{equation}
which are free parameters because the full theory is unknown.
One should stress that the usefulness of any EFT analysis of a given process relies on the assumption that only few terms in the expansion of Eq.~(\ref{lagrangian}) give for that process an adequate approximation to the underlying UV theory.  
The necessary condition obviously is that the energy scale
of the considered process, $E<\Lambda$.
%\footnote{\color{red} For $WW$ scattering, we consider  large center of mass energy and large scattering angles, namely large Mandelstam variables $s\sim t\sim u\sim E^2 \gg M^2_W$. }
However, the effective parameters in the expansion 
Eq.~(\ref{lagrangian}) are the $f$'s and not the scale $\Lambda$ itself.  Neither 
$\Lambda$ nor the $C$'s are known without referring to specific UV complete models. 
Even for $E<<\Lambda$ a simple counting of powers of $E/\Lambda$ can be misleading as far
as the contribution of various operators to a given process is concerned.  The latter 
depends also on the relative magnitude of the couplings $C$, e.g., $C_i^{(6)}$ versus 
$C_i^{(8)}$ and/or within each of those sets of operators, separately, \cite{Giudice:2007fh,Biekoetter:2014jwa,Liu:2016idz,
Contino:2016jqw}, as well as on the interference patterns 
in various amplitudes
calculated from the Lagrangian Eq.~(\ref{lagrangian}) \cite{Azatov:2016sqh}.

For instance, the contribution of dimension-6 (D=6) operators to a given process can be 
suppressed compared to dimension-8 (D=8) operators contrary to a naive 
$(E/\Lambda)$ power counting \cite{Liu:2016idz,Contino:2016jqw,Azatov:2016sqh, 
Franceschini:2017xkh} or, vice versa, the
[SM $\times$ D=8] interference contribution can be subleading with respect to the 
$[D=6]^2$ one
\cite{Biekoetter:2014jwa, Contino:2016jqw,Falkowski:2016cxu}.  Clearly, the assumption about the choice of 
operators in the truncation in Eq.~(\ref{lagrangian}) used to analyze a process of our 
interest introduces a strong model dependent aspect of that analysis: one is implicitly 
assuming that there exist a class of UV complete models such that the chosen truncation
is a good approximation.  It is convenient to introduce the concept of EFT ``models" defined by the choice of operators ${\cal O}_i$ and the values of $f_i$.  The question of this paper is then about the discovery potential at the LHC for BSM physics described by various EFT ``models".

The crucial question is what the range of validity can be of a given EFT ``model".  
There is no precise answer to this question unless one starts with
a specific theory and derives Eq.~(\ref{lagrangian}) by decoupling the new degrees of 
freedom.  However, in addition to the obvious constraint that the EFT approach can be 
valid only for the energy scale $E<\Lambda$ (unfortunately with unknown value 
of $\Lambda$), for theoretical consistency the partial wave amplitudes 
should satisfy the perturbative unitarity condition.  The latter
requirement translates into the condition $E^2<\Lambda^2\leq s^U$, where $s^U\equiv s^U(f_i)$ is the 
perturbative partial wave unitarity bound as a function of the chosen operators and the 
values of the coefficients $f_i$'s. 
Thus, the value of $\Lambda^2_{max}=s^U$ gives the upper bound on the validity of the
EFT based ``model". 
Since the magnitude  of the expected (or observed) experimental effects also depends on
the same $f_i$, one has a frame for a consistent use of the EFT ``model" to describe
the data once they are available.
For a BSM discovery in the EFT framework, proper usage of the ``model" is a vital issue.
It makes no physical sense to extend the EFT ``model" beyond its range of applicability,
set by the condition $E<\Lambda$.
We shall illustrate this logic in more detail in the following.

A common practice in the LHC data analyses in the EFT framework is to derive uncorrelated 
limits on one operator at a time while setting all the remaining Wilson coefficients
to zero.  This in fact means {\it choosing} different EFT ``models":
such limits are valid only under the assumption that just
one chosen operator dominates BSM effects in the studied process in the available energy
range.  In this paper we will consider only variations of single dimension-8 operators
\footnote{ For a physical justification of omitting dimension-6 operators
see Section 2.}.
However, the strategy we present can be extended to the case of many operators at a time,
including dimension-6
(keeping in mind that varying more than one operator substantially
complicates the analysis).
For a given EFT ``model"
\begin{equation}
\sigma \propto |A_{full}|^2=|A_{SM}|^2+(A_{SM}\times A_{BSM}^*+hc)+|A_{BSM}|^2.
\label{full1}
\end{equation}

We focus on the process
\begin{equation}
pp\rightarrow 2 jets + W^+ W^+ \rightarrow 2 jets + l^+\nu +l'^+\nu^\prime
\label{process}
\end{equation}
where $l$ and $l'$ stand for any combination of electrons and muons.
The process depends on the $W^+W^+$ scattering amplitude (the gauge bosons can of course be virtual).  
The EFT ``models" can be maximally valid up to certain invariant mass $M=\sqrt{s}$ of the 
$W^+W^+$ system 
\begin{equation}
M<\Lambda\leq M^U(f_i)
\label{cutoff}
\end{equation}
where $M^U(f_i)$ is fixed by the partial wave perturbative unitarity constraint, $(M^U(f_i))^2=s^U(f_i)$.

The differential cross section
$\frac{d\sigma}{dM}$ reads (actual 
calculations must include also all non-VBS diagrams leading to the same final states):
\begin{equation}
\frac{d\sigma}{dM}\sim \Sigma_{ijkl} \int dx_1 dx_2 q_i(x_1) q_j(x_2) | {\cal M}(ij\to klW^+W^+)|^2 d\Omega\; 
\delta(M-\sqrt{(p_{W^+}+p_{W^+})^2})
\label{cross}
\end{equation}
where $q_i(x)$ is the PDF for parton $i$, the sum runs over partons in the initial ($ij$)
and final ($kl$) states and over helicities, the amplitude ${\cal M}$ is for the parton level process $ij\to
klW^+W^+$ and $d\Omega$ denotes the final state phase space integration.  The special 
role of the distribution $\frac{d\sigma}{dM}$ follows from the fact that it is 
straightforward to impose the cutoff $M\leq\Lambda$, Eq.~(\ref{cutoff}), for the $WW$ scattering amplitude. 
The differential cross section, $\frac{d\sigma}{dM}$, is therefore a very sensitive and straightforward test of new physics defined by a given EFT ``model". Unfortunately, the $W^+W^+$ invariant 
mass in the purely leptonic $W$ decay channel is not directly accessible experimentally 
and one has to investigate various experimental distributions of the charged particles.
The problem here is that the kinematic range of those distributions is not related to the EFT ``model" validity cutoff $M<\Lambda$ and if $\Lambda < M_{max}$, where $M_{max}$ is the 
kinematic limit accessible at the LHC for the $WW$ system, there is necessarily also a 
contribution to those distributions from the region $\Lambda < M < M_{max}$.  
The question is then: in case a deviation from SM predictions is indeed observed,
how to verify a ``model" defined by a single higher-dimension operator ${\cal O}_i^{(k)}$
and a given value of $f_i$ by fitting it to a set of experimental distributions $D_i$ 
and in what range of $f_i$ such a fit is really meaningful \cite{Falkowski:2016cxu}. Before we address this question, it is in order to comment on the perturbative partial wave unitarity constraint.

It is worthwhile to stress several interesting points.

\begin{enumerate}
\item For a given EFT ``model", the unitarity bound is very different for the $J=0$ partial 
wave of different helicity amplitudes and depends on their individual energy dependence
(some of them remain even constant and never violate unitarity, see Appendix).
Our $M^U$ has to be taken as the $lowest$ unitarity bound, universally for all helicity 
amplitudes, because it is the lowest bound that determines the scale $\Lambda_{max}$.
More precisely, one should take the value obtained from diagonalization of the matrix of the $J=0$ partial waves in the helicity space.
\item Correct assessment of the EFT ``model" validity range in the $W^+W^+$ scattering process
requires also consideration of the $W^+W^-$ scattering amplitudes which by construction
probe the same couplings and are sensitive to exactly the same operators.  For most
higher dimension operators,
this actually significantly reduces their range of validity in $W^+W^+$ analyses.
Conversely, the $WZ$ and $ZZ$ processes can be assumed to contain uknown
contributions from additional operators which adjust the value of $\Lambda$ consistently.
\item 
It is interesting to note that for the $f_i$ values of practical interest
the deviations 
from SM predictions in the total cross sections become sizable only in a 
narrow range of energies just below the value of $M^U$, where the $|A_{BSM}|^2$ term
in Eq.~(\ref{full1}) takes over.  However, for most dimension-8 operators the contribution
of the interference term is not completely negligible (see Appendix for details).
Even if deviations from the SM are dominated by the helicity combinations that reach the 
unitarity bound first, the total unpolarized cross sections up to
$M=M^{U}$ get important contributions also from amplitudes which are still far from their own
unitarity limits.  
\end{enumerate}
In the Appendix we illustrate various aspects of those bounds by presenting the results of analytical calculations for two dimension-8 operators, one contributing mainly to the scattering of longitudinally polarized gauge bosons and one to transversely polarized.

We now come back to the problem of testing the EFT ``models" when the $W^+W^+$ invariant mass is not accesible experimentally.
Let us define the BSM signal as the deviation from the SM prediction in the distribution
of some observable $D_i$.
\begin{equation}
S=D_i^{model}-D_i^{SM}.
\label{N}
\end{equation}
The first quantitative estimate of the signal can be written as
\begin{equation}
D_i^{model}=\int^{\Lambda}_{2M_W}\frac{d\sigma}{dM}|_{model} dM +\int_{\Lambda}^{M_{max}}\frac{d\sigma}{dM}|_{SM} dM.
\label{dsigma}
\end{equation}
It defines signal coming uniquely from the operator that defines the ``model" in its range
of validity and assumes only the SM contribution in the region $M>\Lambda$.
Realistically one expects some BSM contribution also from the region above
$\Lambda$.  While this additional contribution may enhance the signal and thus
our sensitivity to new physics, it may also preclude proper description of the data in the EFT 
language.  Such description in terms of a particular EFT ``model" makes sense if and only if this contribution is small enough  when compared to the contribution from the region controlled by the EFT ``model".  The latter depends on the value of $\Lambda$ and $f_i$, and the former on the unknown physics for $M>\Lambda$, which regularizes the scattering amplitudes and makes them consistent with partial wave unitarity.
Ideally, one would conclude that the EFT ``model" is tested for values of $(\Lambda\leq M^U, f_i)$ 
such that the signals computed from Eq.~(\ref{dsigma}) are statistically consistent 
(say, within 2 standard deviations) with the signals computed when the tail $\Lambda>M^U$ is 
modeled in {\it any} way that preserves unitarity of the amplitudes,
i.e., the contribution from this region is sufficiently suppressed kinematically by
parton distributions.  This requirement is of course impossible to impose in practice, but
for a rough quantitative estimate of the magnitude of this 
contribution, one can assume that all the helicity amplitudes above 
$\Lambda$ remain 
constant at their respective values they reach at $\Lambda$, and that $\Lambda$ is 
common to all the helicity amplitudes.
For $\Lambda = \Lambda_{max}$, this prescription regularizes the helicity amplitudes that violate unitarity at $M^U$ and also properly accounts for the contributions of the helicity amplitudes that remain constant with energy. It gives a reasonable approximation to the total unpolarized cross sections for $M>M^U$, at least after some averaging over $M$.
More elaborated regularization techniques can also be checked  here.
The full contribution to a given distribution $D_i$ is then taken as

\begin{equation}
D_i^{model}=\int^{\Lambda}_{2M_W}\frac{d\sigma}{dM}|_{model} dM +\int_{\Lambda}^{M_{max}}\frac{d\sigma}{dM}|_{A=const} dM
\label{unitarized}
\end{equation}

BSM observability imposes some minimum value of $f$ to obtain the required
signal statistical significance.
It can be derived based on Eq.~(\ref{unitarized}) (or Eq.~(\ref{dsigma})).  On the other
hand, description in the EFT language imposes some maximum value of $f$ such that signal
estimates computed from Eqs.~(\ref{dsigma}) and (\ref{unitarized}) remain 
statistically consistent.  Large difference between the two computations implies
significant sensitivity to the region above $\Lambda$.  It impedes a meaningful data
description in the EFT language and also suggests we are more likely to observe
the new physics directly.

\begin{figure}[hbtp]
\centering
\vspace{-2.5cm}
\includegraphics[width=1.0\linewidth]{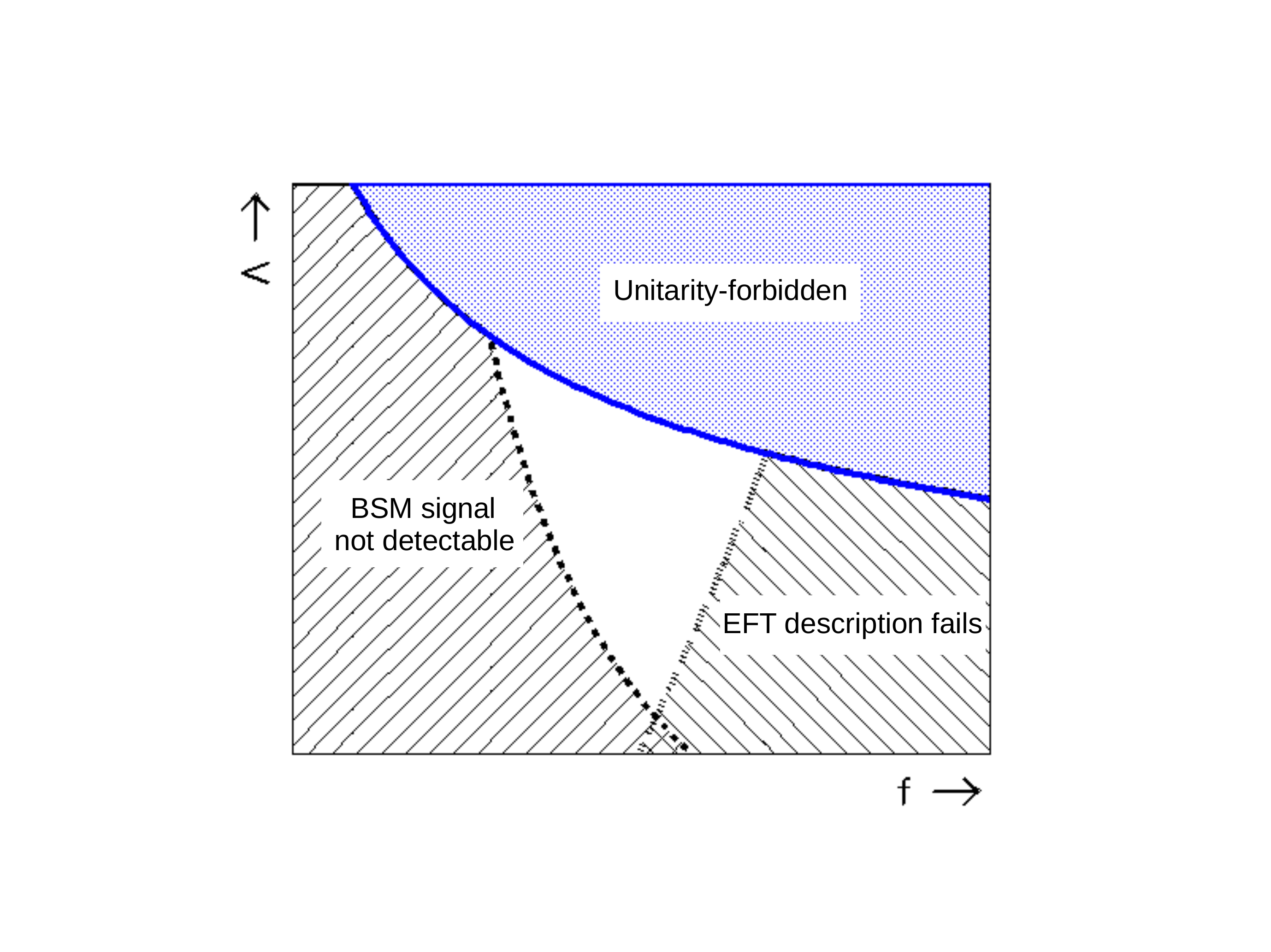}
\vspace{-2.5cm}
\caption{
Cartoon plot which shows the regions in $f_i$ and $\Lambda$ (for an arbitrary
higher-dimension operator ${\cal O}_i$) in terms of BSM signal observability
and applicability of EFT ``models" based on the choice of a higher-dimension operator in an analysis of the same-sign VBS process with purely 
leptonic decays.  The central white triangle is the most interesting region
where the underlying BSM physics can be studied within the EFT framework.
}
%\vspace{1cm}
\label{fig:cartoonplot}
\end{figure}

Assuming $\Lambda=M^U$, we get a finite interval of possible $f$ values, bounded from two sides,
for which BSM discovery and correct EFT description are both plausible.
In the more general case when
$\Lambda <M^U$, i.e., new physics states may appear before our EFT ``model" reaches its 
unitarity limit, respective limits on $f$  depend on the actual value of $\Lambda$.
We thus obtain a 2-dimensional region in the plane $(\Lambda, f_i)$, which is
shown in the cartoon plot in Fig.~\ref{fig:cartoonplot}.  This region is bounded from 
above by the unitarity bound $M^U(f_i)$ (solid blue curve), from the left by the
signal significance criterion (dashed black curve)
and from the right by the EFT consistency criterion (dotted black curve).
The EFT could be the right framework to search for BSM physics as long as these
three criteria do not mutually exclude each other, i.e., graphically, the ``triangle"
shown in our cartoon plot is not empty.  In Section 3 we will verify whether such 
``triangles" indeed exist for the individual dimension-8 operators.

Thus, our preferred strategy for data analysis is as follows:

\begin{enumerate}
\item From collected data measure a distribution $D_i$ (possibly in more than one
dimension) that
offers the highest sensitivity to the studied operator(s),
\item If deviations from the SM are indeed observed \footnote{ We do not discuss in this paper the bounds on the Wilson coefficients obtained from the data analysis when no statistically significant signal on new physics is observed. Such an analysis requires a separate discussion, although it will be also influenced by the results of this paper.}, fit particular values of 
$(\Lambda\leq M^U, f_i)$ based on EFT simulated templates in which
the contribution from the region $M > \Lambda$ is taken into account according to 
Eq.~(\ref{unitarized}) or using some more elaborated regularization methods,
\item Fixing $f_i$ and $\Lambda$ to the fit values, recalculate the 
$D_i$ template
so that the region $M > \Lambda$ is populated only by the SM contribution
(Eq.~(\ref{dsigma})),
\item Check statistical consistency between the original simulated $D_i$ template
and the one based on Eq.~(\ref{dsigma}),
\item Physics conclusions from the obtained $(\Lambda, f_i)$ values can only be drawn
if such consistency is found.  
In addition, stability of the result against different
regularization methods provides a measure of uncertainty of the procedure - too much
sensitivity to the region above $\Lambda$ means the procedure is destined to fail
and so the physical conclusion is that data cannot be described with the studied operator.
\end{enumerate}

\section{Preliminary technicalities}

The same-sign $pp\to W^+W^+ jj$ process probes a number of higher dimension operators.
Among them are dimension-6 operators which modify only the Higgs-to-gauge coupling:
\begin{equation}
\begin{aligned}
{\cal O}_{\Phi d} = \partial_\mu(\Phi^\dagger\Phi)\partial^\mu(\Phi^\dagger\Phi),\\
{\cal O}_{\Phi W} = (\Phi^\dagger\Phi)\mbox{Tr}[W^{\mu\nu}W_{\mu\nu}],\\
{\cal O}_{\tilde WW}=\Phi^\dagger{\tilde W}_{\mu\nu}W^{\mu\nu}\Phi
\end{aligned}
\end{equation}
(the last one being CP-violating),
dimension-6 operators which induce anomalous triple gauge couplings (aTGC):
\begin{equation}
\begin{aligned}
{\cal O}_{WWW}=\mbox{Tr}[W_{\mu\nu}W^{\nu\rho}W_{\rho}^{\mu}],\\
{\cal O}_W=(D_\mu\Phi)^\dagger W^{\mu\nu}(D_\nu\Phi),\\
{\cal O}_B=(D_\mu\Phi)^\dagger B^{\mu\nu}(D_\nu\Phi),\\
{\cal O}_{\tilde WWW}=\mbox{Tr}[{\tilde W}_{\mu\nu}W^{\nu\rho}W_{\rho}^{\mu}],\\
{\cal O}_{\tilde W}=(D_\mu\Phi)^\dagger {\tilde W}^{\mu\nu}(D_\nu\Phi)
\end{aligned}
\end{equation}
(the last two of which are CP-violating),
as well as dimension-8 operators which induce only anomalous quartic couplings (aQGC).
%that modify the interactions among electroweak gauge bosons which are relevant for this process
%\begin{equation}
% \begin{aligned}
%   {\cal O}_{WWW}&=\mbox{Tr}[W_{\mu\nu}W^{\nu\rho}W_{\rho}^{\mu}],\\
%{\cal O}_W&=(D_\mu\Phi)^\dagger W^{\mu\nu}(D_\nu\Phi),\\
%{\cal O}_B&=(D_\mu\Phi)^\dagger B^{\mu\nu}(D_\nu\Phi),\\
%{\cal O}_{\tilde WWW}&=\mbox{Tr}[{\tilde W}_{\mu\nu}W^{\nu\rho}W_{\rho}^{\mu}],\\
%{\cal O}_{\tilde W}&=(D_\mu\Phi)^\dagger {\tilde W}^{\mu\nu}(D_\nu\Phi),
%\label{eq:dim6}
%\end{aligned}
%\end{equation}
In the above,
$\Phi$ is the Higgs doublet field, 
the covariant derivative is defined as
\begin{equation}
D_\mu \equiv \partial_\mu + i \frac{g'}{2} B_\mu  + i g W_\mu^i \frac{\tau^i}{2}
\end{equation}
and the field strength tensors are
\begin{equation}
\begin{aligned}
 W_{\mu\nu} =  \frac{i}{2} g\tau^i (\partial_\mu W^i_\nu - \partial_\nu W^i_\mu
       + g \epsilon_{ijk} W^j_\mu W^k_\nu ), 
\\
 B_{\mu \nu}  = \frac{i}{2} g' (\partial_\mu B_\nu - \partial_\nu B_\mu)
\end{aligned}
\label{eq:fields}
\end{equation}
for gauge fields $W^i_\mu$ and $B_\mu$ of $SU(2)_I$ and $U(1)_Y$, respectively.  

Higgs and triple gauge couplings can be accessed experimentally via other processes, 
namely Higgs physics and diboson production which is
most sensitive to aTGC.  They are presently known to agree with the SM within a
few per cent \cite{Butter:2016cvz}, which translates into stringent limits on the
dimension-6 operators.

On the other hand, VBS processes are more suitable to constrain aQGC.
The following dimension-8 operators contribute to the $WWWW$ vertex:
\begin{equation}
\begin{aligned}
  {\cal O}_{S0} &= \left [ \left ( D_\mu \Phi \right)^\dagger
 D_\nu \Phi \right ] \times
\left [ \left ( D^\mu \Phi \right)^\dagger
D^\nu \Phi \right ],
\\
  {\cal O}_{S1} &= \left [ \left ( D_\mu \Phi \right)^\dagger
 D^\mu \Phi  \right ] \times
\left [ \left ( D_\nu \Phi \right)^\dagger
D^\nu \Phi \right ],
\\
 {\cal O}_{M0} &=   \hbox{Tr}\left [ \hat{W}_{\mu\nu} \hat{W}^{\mu\nu} \right ]
\times  \left [ \left ( D_\beta \Phi \right)^\dagger
D^\beta \Phi \right ],
\\
 {\cal O}_{M1} &=   \hbox{Tr}\left [ \hat{W}_{\mu\nu} \hat{W}^{\nu\beta} \right ]
\times  \left [ \left ( D_\beta \Phi \right)^\dagger
D^\mu \Phi \right ],
\\
  {\cal O}_{M6} &= \left [ \left ( D_\mu \Phi \right)^\dagger \hat{W}_{\beta\nu}
\hat{W}^{\beta\nu} D^\mu \Phi  \right ],
\\
  {\cal O}_{M7} &= \left [ \left ( D_\mu \Phi \right)^\dagger \hat{W}_{\beta\nu}
\hat{W}^{\beta\mu} D^\nu \Phi  \right ],
\\
 {\cal O}_{T0} &=   \hbox{Tr}\left [ \hat{W}_{\mu\nu} \hat{W}^{\mu\nu} \right ]
\times   \hbox{Tr}\left [ \hat{W}_{\alpha\beta} \hat{W}^{\alpha\beta} \right ],
\\
 {\cal O}_{T1} &=   \hbox{Tr}\left [ \hat{W}_{\alpha\nu} \hat{W}^{\mu\beta} \right ] 
\times   \hbox{Tr}\left [ \hat{W}_{\mu\beta} \hat{W}^{\alpha\nu} \right ],
\\
 {\cal O}_{T2} &=   \hbox{Tr}\left [ \hat{W}_{\alpha\mu} \hat{W}^{\mu\beta} \right ]
\times   \hbox{Tr}\left [ \hat{W}_{\beta\nu} \hat{W}^{\nu\alpha} \right ] .
\end{aligned}
\end{equation}

In the above, we have defined $\hat{W}_{\mu\nu} = \frac{1}{ig}W_{\mu\nu}$.
Throughout this paper we follow the convention used in MadGraph \cite{MadGraph} with
dimension-8 operators included via public UFO files as far as
the actual definitions of the field strength tensors and
Wilson coefficients are 
concerned.  Whenever results from the VBFNLO program \cite{VBFNLO} are used in this work,
appropriate conversion factors are applied.  For more details on the subject see
Ref.~\cite{Degrande:2013rea}.

The same-sign $pp\to W^+W^+ jj$ production has been already observed during Run I of 
the LHC \cite{Aad:2014zda,WWrun1} and confirmed
by a recent measurement of the CMS Collaboration at 13 TeV Run II \cite{Sirunyan:2017ret}.
Also, pioneering measurements of the $ZW^\pm jj$ \cite{Aad:2016ett} and $ZZ jj$ 
\cite{cms_vbs_zz} processes
exist.  They all place experimental limits on the relevant dimension-8 operators.
However, most presently obtained limits involve unitarity violation within the
measured kinematic range, leading to problems in physical interpretation and even
comparison of the different analyses.

Our goal is to investigate the discovery potential at 
the High Luminosity LHC (HL-LHC) of the BSM physics effectively described by EFT ``models" 
with single 
dimension-8 operators at a time, with proper attention paid to the regions of validity of such models, as described in Section 1.

\section{Results of simulations}

For the following analysis dedicated event samples of the process 
$pp \to jj\mu^+\mu^+\nu\nu$ at 14 TeV
were generated at LO using the MadGraph5\_aMC@NLO v5.2.2.3 generator \cite{MadGraph}, 
with the appropriate UFO files 
containing additional vertices involving the desired dimension-8 operators.
For each dimension-8 operator a sample of at least 500,000 events within a phase space 
consistent with a VBS-like topology (defined below) was generated.  A preselected arbitrary value of 
the relevant $f$ coefficient (from now on, $f \equiv f_i$ with $i = S0, S1, T0, T1, T2,
M0, M1, M6, M7$) was assumed at each generation;
different $f$ values were obtained by applying weights to generated events, using
the reweight command in MadGraph.  The value $f$=0 represents the Standard 
Model predictions for each study.
The Pythia package v6.4.1.9 \cite{Pythia6} was used for hadronization as well as initial and final state 
radiation processes.  No detector was simulated.  Cross sections at the
output of MadGraph were multiplied by a factor 4 to
account for all the lepton (electron and/or muon) combinations in the final state.

In this analysis, the Standard Model process $pp \to jjl^+l^+\nu\nu$ is treated
as the irreducible background, while signal is defined as the enhancement
(which may be positive or negative in particular cases) of the event
yield in the presence of a given dimension-8 operator relative to the
Standard Model prediction.  No reducible backgrounds were simulated, as they
are known to be strongly detector dependent.  For this reason, results presented
here should be treated mainly as a demonstration of our strategy rather than as a precise determination
of numerical values.
For more realistic results this analysis should be repeated with full
detector simulation for each of the LHC experiments separately.

The final analysis is performed by applying standard VBS-like event
selection criteria, similar to those applied in data analyses carried
by ATLAS and
CMS.  These were: $M_{jj} >$ 500 GeV, $\Delta\eta_{jj} >$ 2.5, $p_T^{~j} >$ 30 GeV,
$|\eta_j| <$ 5, $p_T^{~l} >$25 GeV, $|\eta_l| <$ 2.5.
As anticipated in Section 1, signal is calculated in two ways.
First, using Eq.~(\ref{dsigma}), where $\Lambda$ can vary in principle between $2M_W$ and
the appropriate unitarity limit for each chosen value of $f$.
The $M_{WW} > \Lambda$ tail of the distribution is then assumed identical as
in the Standard Model case.
Second, using Eq.~(\ref{unitarized}) which accounts for an additional BSM contribution
coming from the region $M_{WW} > \Lambda$.  The latter is estimated under the
assumption that helicity amplitudes remain constant above this limit, as discussed
in Section 1.  For the case when $\Lambda$ is equal to the
unitarity limit, this corresponds to unitarity saturation.

\begin{figure}[hbtp]
\centering
\includegraphics[width=1.05\linewidth]{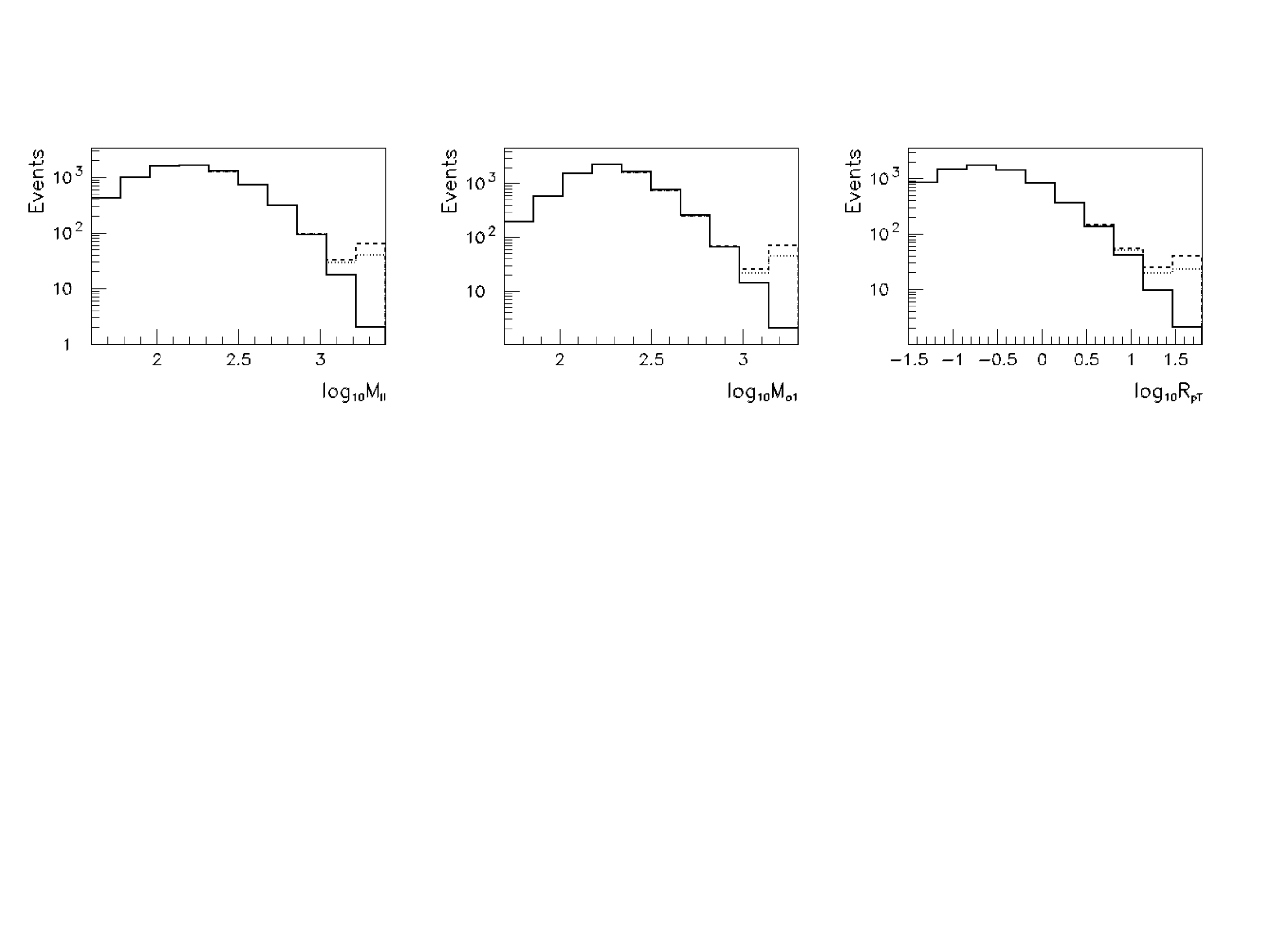}
\vspace{-8.5cm}
\caption{
Typical examples of kinematic distributions used for the assessment of BSM signal significances.
Shown are the distributions of $M_{ll}$, $M_{o1}$ and $R_{pT}$ (in log scale): in the
Standard Model (solid lines), with $f_{T1}$=0.1/TeV$^{-4}$ and the high-$M_{WW}$ tail
treatment according to Eq.~(\ref{unitarized}) (dashed lines), and with
$f_{T1}$=0.1/TeV$^{-4}$ and the high-$M_{WW}$ tail
treatment according to Eq.~(\ref{dsigma}) (dotted lines).
Assumed is $\sqrt{s}$ = 14 TeV and an integrated luminosity of 3 ab$^{-1}$.
}
\label{fig:Dim8Dist}
\end{figure}

For each $f$ value of every dimension-8 operator, signal significance is
assessed by studying the distributions of a large number of kinematic variables.  
We only considered one-dimensional distributions
of single variables.  Each distribution was divided into 10 bins,
arranged so that the Standard Model prediction in each bin is never lower than
2 events.  Overflows were always included in the respective highest bins.
Ultimately, each distribution had the form of 10 numbers, that represent the expected
event yields normalized to a total integrated luminosity of 3 ab$^{-1}$, each calculated in three
different versions: $N^{SM}_i$ for the Standard Model case, $N^{EFT}_i$ from
applying Eq.~(\ref{dsigma}), and $N^{BSM}_i$ from applying Eq.~(\ref{unitarized})
(here subscript $i$ runs over the bins).
In this analysis, Eq.~(\ref{unitarized}) was implemented by applying additional
weights to events above $M_{WW} = \Lambda$ in the original non-regularized
samples generated by MadGraph.  For the dimension-8 operators, this weight
was equal to $(\Lambda/M_{WW})^4$.
The choice of the power in the exponent takes into account that the non-regularized
total cross section for $WW$ scattering grows less steeply around $M_{WW} = \Lambda$
than its asymptotic
behavior $\sim s^3$, which is valid in the limit $M_{WW} \to \infty$.
This follows from the observation that unitarity is first violated much before
the cross section gets dominated by its $\sim s^3$ term, as shown in the Appendix.
The applied procedure is supposed to ensure that the total $WW$ scattering
cross section after regularization behaves like $1/s$ for $M_{WW} > \Lambda$,
and so it approximates the
principle of constant amplitude (Section 1), at least after some averaging over the
individual helicity combinations.
Examples of simulated distributions are shown in Fig.~\ref{fig:Dim8Dist}.

Signal significance expressed in standard deviations ($\sigma$)
is defined as the square root
of a $\chi^2$ resulting from comparing the bin-by-bin event yields:

\begin{equation}
\chi^2 = \sum_i (N^{BSM}_i - N^{SM}_i)^2 / N^{SM}_i.
\end{equation}

\begin{figure}[hbtp]
\centering
\includegraphics[width=0.95\linewidth]{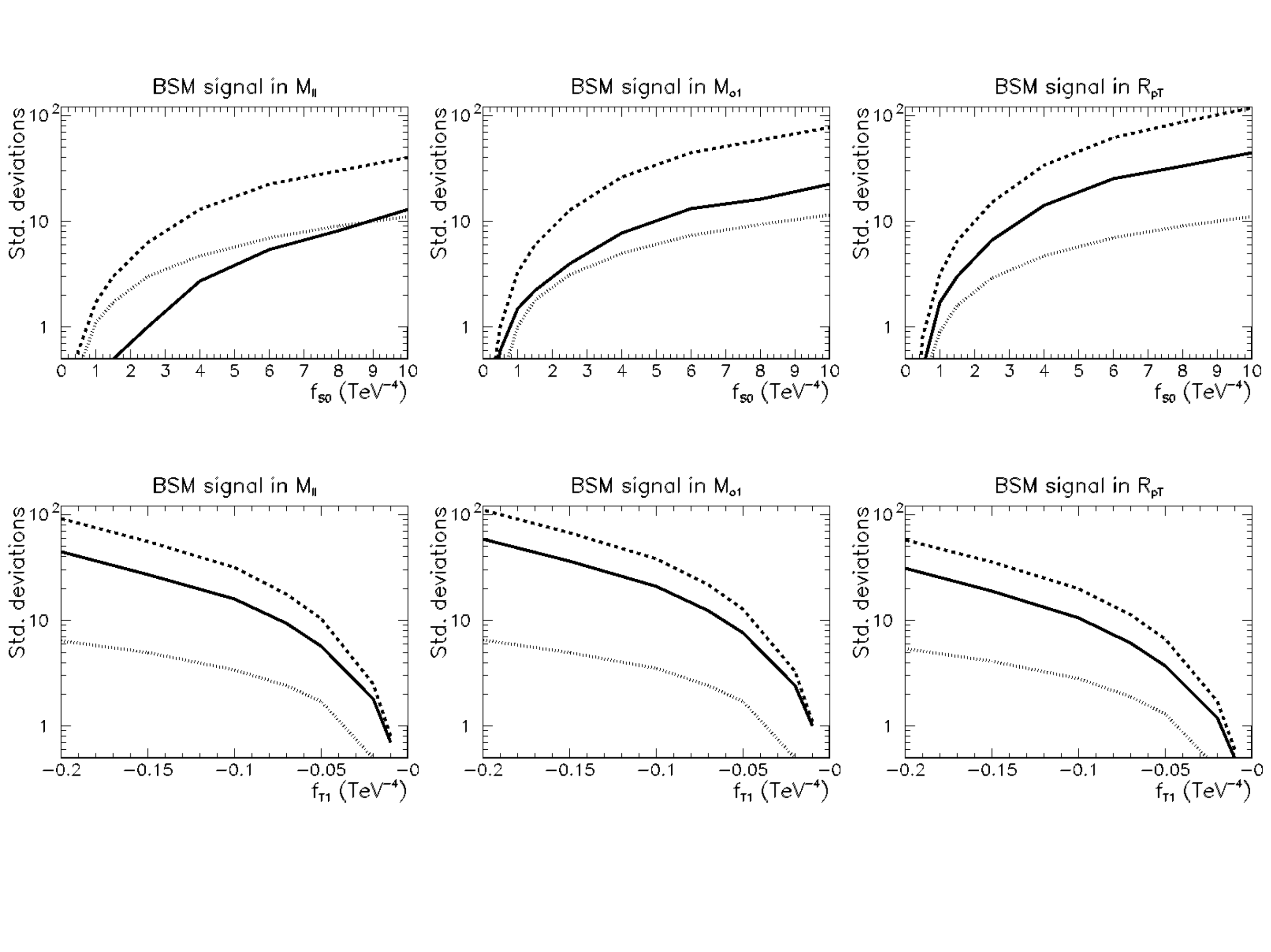}
\vspace{-1.5cm}
\caption{
Typical examples of BSM signal significances computed as a function of $f_{S0}$ (upper row)
and $f_{T1}$ (lower row) based on different kinematic distributions.  Here the $\Lambda$
cutoff is assumed equal to the unitarity limit.  Shown are predictions obtained by using
Eq.~(\ref{dsigma}) (solid lines) and Eq.~(\ref{unitarized}) (dashed lines).
The dotted lines show the difference in standard deviations between the two respective 
calculations.
Assumed is $\sqrt{s}$ = 14 TeV and an integrated luminosity of 3 ab$^{-1}$.
}
\label{fig:Sigmas}
\end{figure}

Lower observation limits on each operator are defined by the requirement of
signal significance being above the 5$\sigma$ level.
Small differences between the respective signal
predictions obtained using Eqs.~(\ref{dsigma}) and (\ref{unitarized}), as well
as using other regularization techniques, will be manifest
as slightly different observation limits
and should be understood as the uncertainty margin
arising from the unknown physics above $\Lambda$, no longer described in terms of 
the EFT.  Examples of signal significances 
as a function of $f$ are shown in
Fig.~\ref{fig:Sigmas} with dashed curves.
Consistency of the EFT description is determined by requiring
a small difference between the respective predictions from 
Eqs.~(\ref{dsigma}) and (\ref{unitarized}).  
An additional $\chi^2_{add}$ is computed based on the comparison of the
respective distributions of $N^{EFT}_i$ and $N^{BSM}_i$:

\begin{equation}
\chi^2_{add} = \sum_i (N^{EFT}_i - N^{BSM}_i)^2 / N^{BSM}_i.
\end{equation}

In this analysis
we allowed differences amounting to up to 2$\sigma$ in the most sensitive
kinematic distribution.  This difference as a function of $f$ is shown in
Fig.~\ref{fig:Sigmas} as dotted curves.
These considerations consequently translate into effective upper limits
on the value of $f$ for each operator.

%This number quantifies our experimental
%sensitivity to the unknown region $M_{WW} > \Lambda$.

For each dimension-8 operator we took the distribution that produced the
highest $\chi^2$ among the considered variables.  The most sensitive
variables we found to be $R_{p_T} \equiv p_T^{~l1}p_T^{~l2}/(p_T^{~j1}p_T^{~j2})$ 
\cite{NaszPapier}
for ${\cal O}_{S0}$ and ${\cal O}_{S1}$, and $M_{o1} \equiv \sqrt{(|\vec{p}_T^{~l1}|+|\vec{p}_T^{~l2}|
+|\vec{p}_T^{~miss}|)^2 - (\vec{p}_T^{~l1}+\vec{p}_T^{~l2}+\vec{p}_T^{~miss})^2}$
\cite{STodt}
for the remaining operators (for some of them, $M_{ll}$ would give almost
identical results as $M_{o1}$, but usually this was not the case).

Unitarity limits were computed using the VBFNLO \cite{VBFNLO} calculator v1.3.0,
after applying appropriate conversion factors to the input values of the Wilson
coeeficients, so to make it suitable to the MadGraph 5 convention.
We used the respective values from T-matrix diagonalization, considering both $W^+W^+$
and $W^+W^-$ channels, and taking always the lower value of the two.
For the operators we consider here, unitarity limits are lower for $W^+W^-$ than 
for $W^+W^+$ except for
$f_{S0}$ (both positive and negative) and negative $f_{T1}$.

\vspace{5mm}

Assuming $\Lambda$ is equal to the respective unitarity bounds,
the lower and upper limits for the values of $f$ for each dimension-8 operator, for positive
and negative $f$ values, estimated for the HL-LHC with an integrated luminosity of 3 ab$^{-1}$,
are read out directly from graphs such as Fig.~\ref{fig:Sigmas} and
listed below in Table 1.  These limits define the (continous) sets of testable 
EFT ``models" based on the choice of single dimension-8 operators.

%\begin{tabular}{c|ccc|c|ccc}
%Operator & Lower limit & Lower limit & Upper limit & Operator & Lower limit & Lower limit & Upper limit \\
%         &  from (\ref{unitarized}) &  from (\ref{dsigma}) & & & from (\ref{unitarized}) &  from (\ref{dsigma}) & \\
%\hline
%$f_{S0}$  & 1.3  & 2.2  & 2.0 & $-f_{S0}$ & 1.2  & 1.8  & 2.0 \\
%$f_{S1}$  & 8.0  & 12   & 6.5 & $-f_{S1}$ & 5.5  & 11   & 6.0 \\ 
%$f_{T0}$  & 0.08 & 0.20 & 0.13 & $-f_{T0}$ & 0.05 & 0.09 & 0.12 \\
%$f_{T1}$  & 0.03 & 0.04 & 0.06 & $-f_{T1}$ & 0.03 & 0.04 & 0.06 \\
%$f_{T2}$  & 0.20 & 0.25 & 0.25 & $-f_{T2}$ & 0.10 & 0.12 & 0.20 \\
%$f_{M0}$  & 1.0  & 1.5  & 1.2 & $-f_{M0}$ & 1.0  & 1.5  & 1.2 \\ 
%$f_{M1}$  & 1.0  & 2.0  & 1.9 & $-f_{M1}$ & 0.9  & 1.2  & 1.8 \\ 
%$f_{M6}$  & 2.0  & 3.0  & 2.4 & $-f_{M6}$ & 2.0  & 3.0  & 2.4 \\
%$f_{M7}$  & 1.1  & 1.4  & 2.8 & $-f_{M7}$ & 1.3  & 2.0  & 2.8 \\
%\end{tabular}
\begin{table}
\vspace{8mm}
\begin{center}
\begin{tabular}{|c|cc|c|cc|}
\hline
Coeff.   & Lower limit & Upper limit & Coeff.   & Lower limit & Upper limit \\
         &  (TeV$^{-4}$) &  (TeV$^{-4}$) & & (TeV$^{-4}$) &  (TeV$^{-4}$) \\
\hline
$f_{S0}$  & 1.3  & 2.0 & $-f_{S0}$ & 1.2  & 2.0 \\
$f_{S1}$  & 8.0  & 6.5 & $-f_{S1}$ & 5.5  & 6.0 \\
$f_{T0}$  & 0.08 & 0.13 & $-f_{T0}$ & 0.05 & 0.12 \\
$f_{T1}$  & 0.03 & 0.06 & $-f_{T1}$ & 0.03 & 0.06 \\
$f_{T2}$  & 0.20 & 0.25 & $-f_{T2}$ & 0.10 & 0.20 \\
$f_{M0}$  & 1.0  & 1.2 & $-f_{M0}$ & 1.0  & 1.2 \\
$f_{M1}$  & 1.0  & 1.9 & $-f_{M1}$ & 0.9  & 1.8 \\
$f_{M6}$  & 2.0  & 2.4 & $-f_{M6}$ & 2.0  & 2.4 \\
$f_{M7}$  & 1.1  & 2.8 & $-f_{M7}$ & 1.3  & 2.8 \\
\hline
\end{tabular}
\end{center}
\vspace{3mm}
\caption{
Estimated lower limits for BSM signal significance and upper limits for
EFT consistency for each dimension-8 operator
(positive and negative $f$ values), for the case when $\Lambda$ is equal
to the unitarity bound,
in the $W^+W^+$ scattering process at the LHC with 3 ab$^{-1}$.}
\end{table}

The fact that the obtained lower limits are more optimistic than those from
several earlier studies (see, e.g., Ref.~\cite{snowmass2013})
reflects our lack of detector simulation and reducible
background treatment, but may be partly due to the use of
the most sensitive kinematic variables.  It must be stressed, nonetheless, that
both these factors affect all lower and upper limits likewise, so their relative
positions with respect to each other are unlikely to change much.

As can be seen, the ranges are rather narrow, but in most cases non-empty.
Rather wide regions where BSM signal significance does not preclude consistent 
EFT description
can be identified for $f_{T1}$ and $f_{M7}$ regardless of sign, as well as
somewhat smaller regions for $f_{T0}$, $f_{T2}$ and $f_{M1}$.
Prospects for $f_{M0}$, $f_{M6}$ and $f_{S0}$ may depend on the accuracy
of the high-$M_{WW}$ tail modeling and
a narrow window is also likely to open up unless measured
signal turns out very close to its most conservative prediction.
Only for positive values of $f_{S1}$, the resulting upper limit for consistent
EFT description remains entirely below the lower limit for signal significance.

\begin{figure}[hbtp]
\centering
\vspace{5mm}
\includegraphics[width=1.02\linewidth]{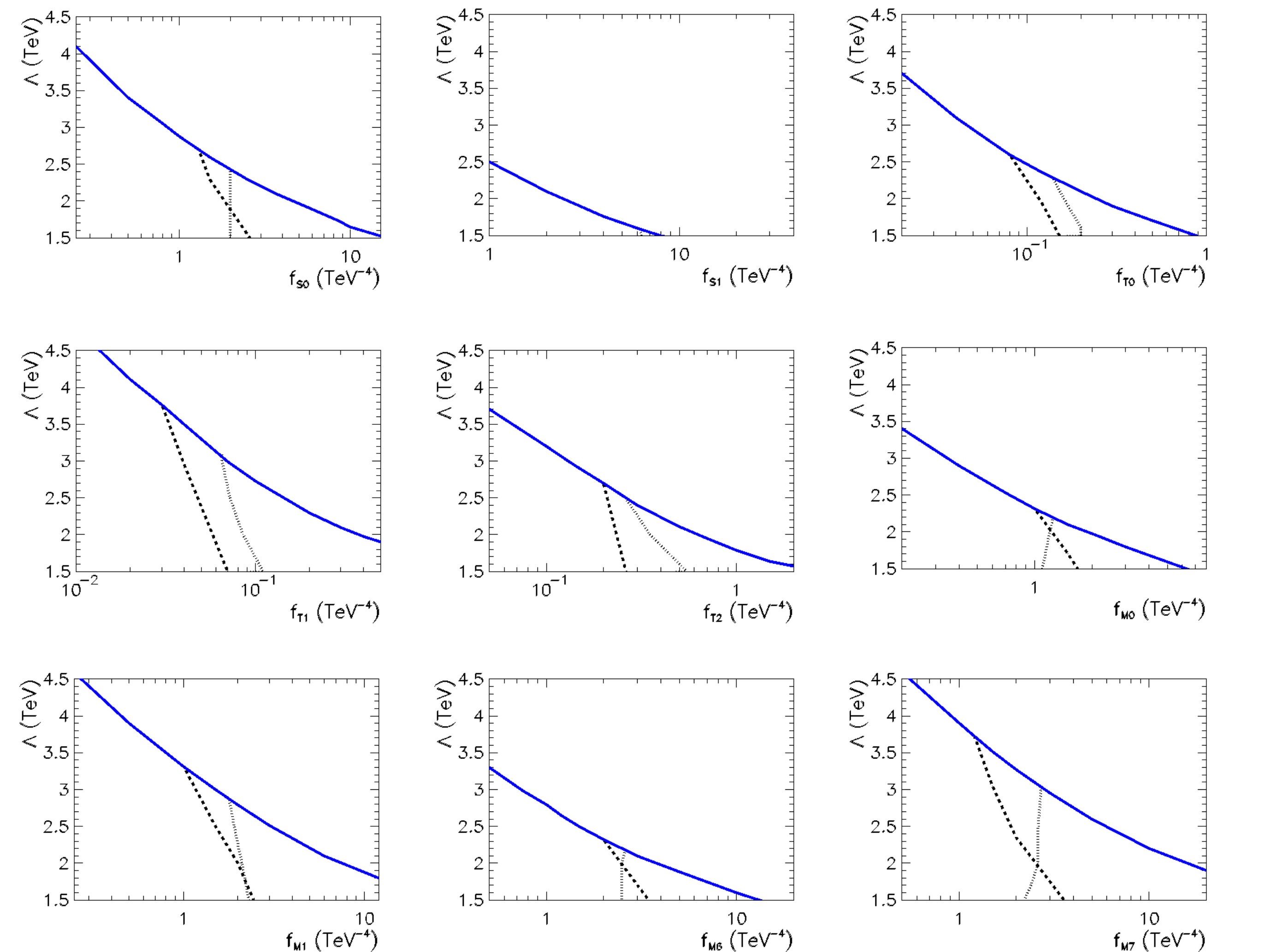}
\vspace{5mm}
\caption{
Regions in the $\Lambda$ vs $f$ (positive $f$ values) space for dimension-8 operators 
in which a 5$\sigma$ BSM signal can be observed and the EFT is applicable.  
The unitarity limit is shown in blue.  Also shown are the lower limits for a 
5$\sigma$ signal
significance from Eq.~(\ref{unitarized}) (dashed lines) and the upper limit on $2\sigma$ EFT
consistency (dotted lines).
Assumed is $\sqrt{s}$ = 14 TeV and an integrated luminosity of 3 ab$^{-1}$.   
}
\label{fig:Dim8Limits3}
\end{figure}

\begin{figure}[hbtp]
\centering
\includegraphics[width=1.02\linewidth]{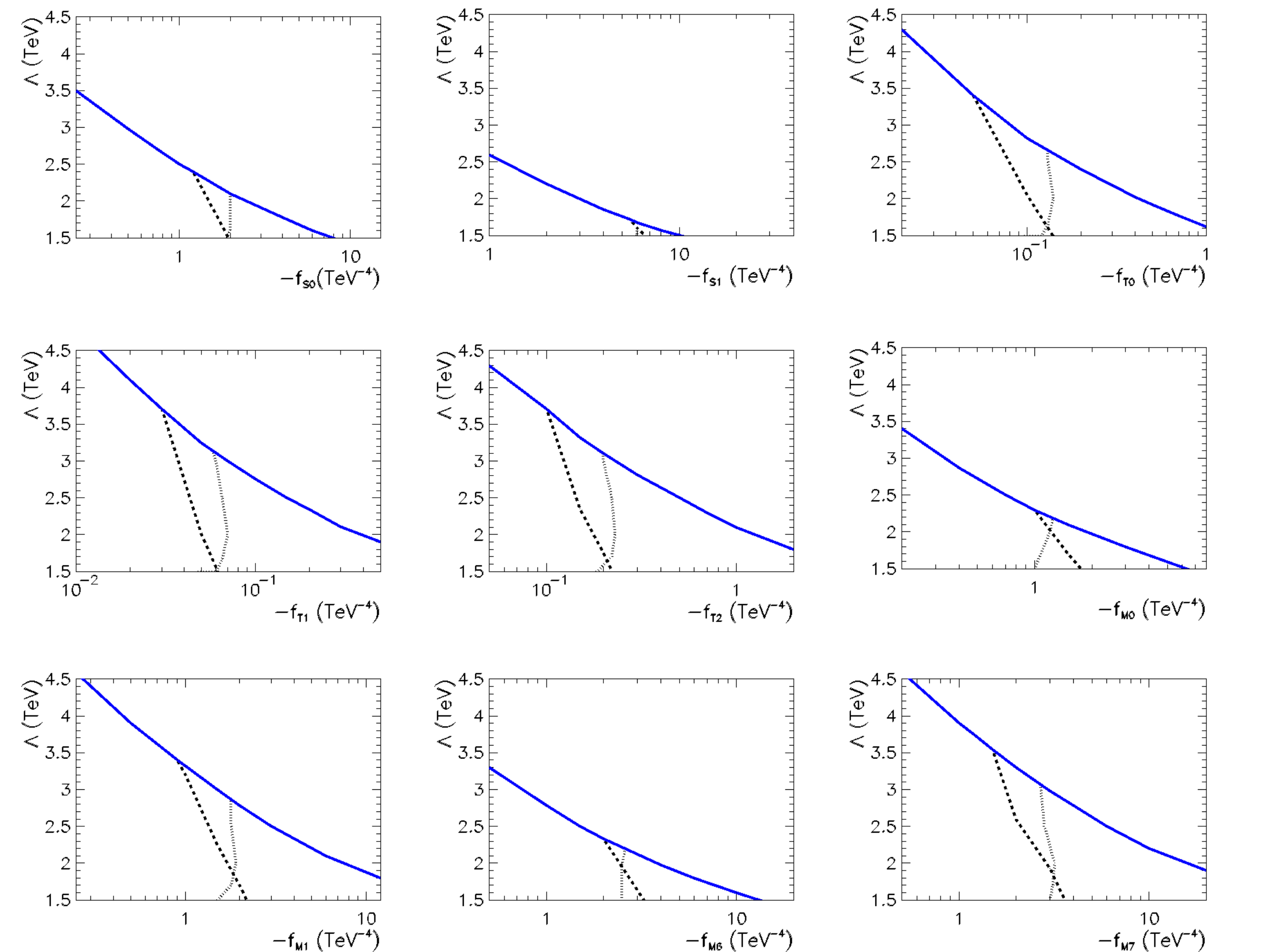}
\vspace{1cm}
\caption{
Regions in the $\Lambda$ vs $f$ (negative $f$ values) space for dimension-8 operators 
in which a 5$\sigma$ BSM signal can be observed and the EFT is applicable.  
For the meaning of curves see caption of Fig.~\ref{fig:Dim8Limits3}.
Assumed is $\sqrt{s}$ = 14 TeV and an integrated luminosity of 3 ab$^{-1}$.        
}
\label{fig:Dim8Limits4}
\end{figure}

Allowing that the scale of new physics $\Lambda$ may be lower than the actual
unitarity bound results in 2-dimensional limits in the $(f,\Lambda)$ plane.
Usually this means further reduction of the allowed $f$ ranges for lower $\Lambda$ 
values and the resulting regions take the form of an irregular triangle.
Respective results for all the dimension-8 operators are depicted in 
Figs.~\ref{fig:Dim8Limits3} and \ref{fig:Dim8Limits4}.  It is interesting to note
that in many cases this puts an effective lower limit on $\Lambda$ itself, in
addition to the upper limit derived from the unitarity condition.
In particular, the adopted criteria bound the value of $\Lambda$ to being above
$\sim$2 TeV for the ${\cal O}_M$ operators as well as for ${\cal O}_{S0}$.  
The ${\cal O}_T$ operators still allow a wider range of $\Lambda$.
Unfortunately, there is little we can learn from fitting $f_{S1}$, since
signal observability requires very low $\Lambda$ values, for which the new physics
could probably be detected directly.

\begin{figure}[hbtp]
\vspace{-2cm}
\centering
\includegraphics[width=1.0\linewidth]{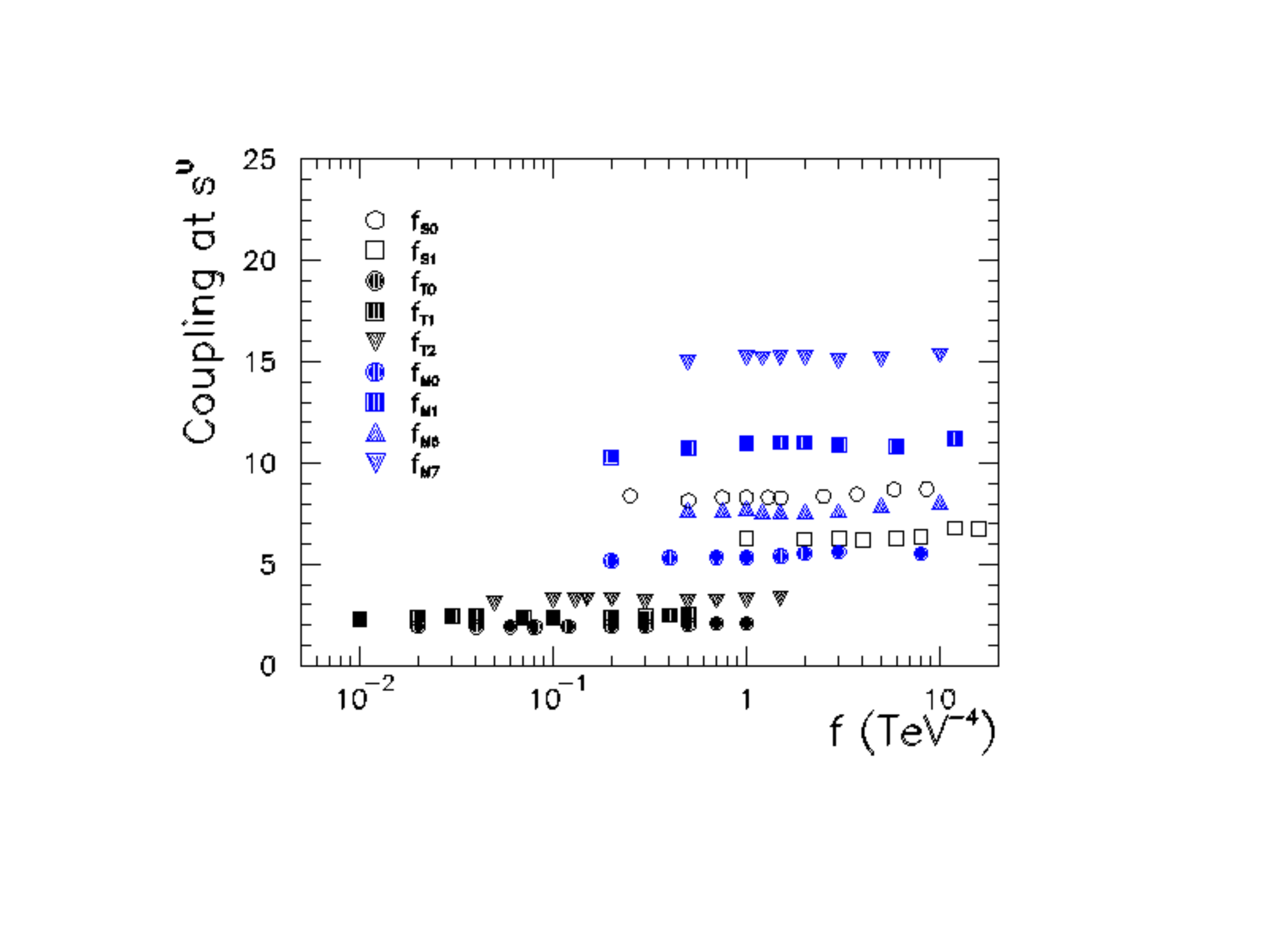}
\vspace{-3cm}
\caption{
Maximum value $\sqrt{C_{max}}$ of the coupling constants related to individual
dimension-8 operators, calculated at the energy where the unitarity limit
is reached, as a function of the relevant $f$ value.
}
\label{fig:slaweksplot}
\end{figure}

It is interesting to plot the values of the couplings $\sqrt{C}$ 
in Eq.~(1.1) as a function of $f_i$  assuming $\Lambda_{max}=M^U$ i.e.,
$C_{max}=f \times (M^{U})^{k-4}$, where $k$ is the dimensionality of the operator that defines the EFT ``model". 
In models with one BSM scale and one BSM coupling constant $\sqrt{C}$ has the interpretation of the coupling constant \cite{Giudice:2007fh}.
The values of $C_{max}$ are to a good approximation
independent of $f$ (see Fig.~\ref{fig:slaweksplot}) and, being generally
in the range ($\sqrt{4\pi}, 4\pi$), reflect the 
approach to a strongly interacting regime in an underlying (unknown) UV complete theory.
The EFT discovery regions depicted in Figs.~\ref{fig:Dim8Limits3} and
\ref{fig:Dim8Limits4} have further interesting 
implications for the couplings $C$.  For a fixed $f$, the unitarity bound $\Lambda^2<s^U$ implies that 
$C<C_{max}=f(M^U)^4$, whereas
the lower bound on $\Lambda$ that comes from the combination of the
signal significance and EFT consistency criteria gives us
$C>\Lambda_{min}^ 4f$.  Thus, a given range $(\Lambda_{min}, \Lambda_{max})$ 
corresponds to a range of values of the couplings $C$, so that we could not only discover
an indirect sign of BSM physics, but also learn something about the nature of the 
complete theory, whether it is strongly or weakly interacting. 
In particular, for the following operators: ${\cal O}_{S0}$, ${\cal O}_{M0}$, ${\cal O}_{M1}$, ${\cal O}_{M6}$ and
${\cal O}_{M7}$, only models with $C$ being close to the strong interaction limit will be
experimentally testable, while a wider range of $C$ may be testable for ${\cal O}_{T0}$,
${\cal O}_{T1}$ and ${\cal O}_{T2}$.

\section{Conclusions and outlook}

In this paper we have analyzed the prospects for discovering physics beyond the SM at the HL-LHC in the EFT framework applied to the VBS amplitudes, in the process $pp\rightarrow W^+W^+jj$.  We have introduced the concept of EFT ``models" defined by the choice of higher dimension operators and values of the Wilson coefficients and analyzed ``models" based on single dimension-8 operators at a time.  
We emphasize the role of the invariant mass $M_{WW}$ whose distribution directly relates
to the intrinsic range of validity of the EFT approach, $M_{WW} < \Lambda\leq M^U$,
and the importance to tackle this issue correctly in data analysis in order to
study the underlying BSM physics.  While this is relatively simple (in principle)
for final states where $M_{WW}$ can be determined on an event-by-event
basis, the value of $M_{WW}$ is unfortunately not available in leptonic $W$ decays.
We argue that usage of EFT ``models" in the analysis of purely
leptonic $W$ decay channels requires bounding the possible contribution from
the region $M_{WW} > \Lambda$, no longer described by the ``model",
and ensuring it does not significantly distort the measured distributions 
compared to what they would have looked from the region of EFT validity alone.

We propose a data analysis strategy to satisfy the above requirements and verify in what
ranges of the relevant Wilson coefficients such strategy can be successfully applied
in a future analysis of the HL-LHC data.  
We find that, with a possible exception of ${\cal O}_{S1}$, all dimension-8 operators which affect the
$WWWW$ quartic coupling have regions where
a 5$\sigma$ BSM signal can be observed at HL-LHC with 3 ab$^{-1}$ of data, while data
could be satisfactorily described using the EFT approach. 

From such analysis it may be possible to learn something about
the underlying UV completion of the full theory.  Successful determination of a
given $f$ value, using a procedure that respects the EFT restricted range of 
applicability,
will put non-trivial bounds on the value of $\Lambda$ and consequently, the BSM
coupling $C$.  These bounds are rather weak for ${\cal O}_{T0}$, ${\cal O}_{T1}$ and ${\cal O}_{T2}$
operators, but potentially stronger for ${\cal O}_{M0}$, ${\cal O}_{M1}$, ${\cal O}_{M6}$ and ${\cal O}_{M7}$.
In particular, applicability of the EFT in terms of these operators already requires
$\Lambda \ge$ 2 TeV, while stringent upper limits arise from the unitarity condition.
Because of relatively low sensitivity to $f_{S0}$ and $f_{S1}$, it will unfortunately
be hard to learn much about $W_LW_L$ physics using the 
EFT approach with dimension-8 operators.

It must be stressed that in this analysis we have only considered single dimension-8
operators at a time.  Allowing non-zero values of more than one $f$ at a time 
provides much more
felixibility as far as the value of $\Lambda$ is concerned, especially for those
operators whose individual unitarity limits are driven by helicity
combinations which contribute little to the total cross section.  Consequently,
regions of BSM observability and EFT consistency can only be larger than what
we found here.  Study of VBS processes in the EFT language can
be the right way to look for new physics and should gain
special attention in case the LHC fails to observe new physics states directly.

Consideration of other VBS processes and $W$ decay channels may significantly 
improve
the situation.  In particular, the semileptonic decays, where one $W^+$ decays
leptonically and the other $W^+$ into hadrons, have never been studied in VBS
analyses because of their more complicated jet combinatorics and consequently much
higher background.  Progress in the implementation of $W$-jet tagging
techniques based on jet substructure algorithms may render these channels interesting
again.
However, they are presently faced with two other experimental challenges.  One is
the precision of the $M_{WW}$ determination which relies on the missing-$E_T$
measurement resolution.  The other one is poor control over the sign over the
hadronic $W$.
The advantages would be substantial.
If $M_{WW}$ can be reconstructed with reasonable accuracy,
it is straightforward to fit $f$ and $\Lambda$ to the measured
distribution in an EFT-consistent way even for arbitrarily large $f$.  Existence
of a high-$M_{WW}$ tail above $\Lambda$ is then not a problem,
but a bonus, as it may give us additional hints about the BSM physics.
Finally, because of the invariant mass issue, the $ZZ$ scattering channel,
despite its lowest cross section, may ultimately prove to be the process from which we
can learn the most about BSM in case the LHC fails to discover new physics directly.

%%%%%%%%%%%%%%%%%%%%%%%%%
\section*{Acknowledgments}
%%%%%%%%%%%%%%%%%%%%%%%%%
The work of SP is partially supported by the National Science Centre, Poland, under research grants
DEC-2014/15/B/ST2/02157, DEC-2015/18/M/ST2/00054 and DEC-2016/23/G/ST2/04301,
and that of JK by the Polish National Science Center HARMONIA project under contract
UMO-2015/18/M/ST2/00518 (2016-2019). JR is supported in part by the National Science Center, Poland, under research grant DEC2015/19/B/ST2/02848.
ST is supported by Fermi Research Alliance, LLC under Contract No.~De-AC02-07CH11359 with the United States Department of Energy.
%%%%%%%%%%%%%%%%%%%%%%%%%

%%%%%%%%%%%%%%%%%%%%%%%%%%%%%%%%%%%%%%%%%%%%

\appendix

\section {Unitarity bounds}

The purpose of this section is to give an overview of the behavior of individual helicity amplitudes as a function of energy and their contributions to the total unpolarized cross section, with special attention paid to the partial wave unitarity constraints, in the SM and in its extensions to the EFT ``models'' discussed in this paper.  We shall illustrate the main points using the operators ${\cal O}_{S0}$ and ${\cal O}_{T1}$.  The choice is determined by the requirement that the $s^U$ bounds for $W^+W^+$ which we present below are stronger than for $W^+W^-$ (often used in our analyses). 
%This is the case for the operator OS0 and for  negative values of $f_i$  fpr OT1.  
The qualitative picture remains the same for the other operators as well.  Analytical computations have been partially performed using a Mathematica code.  All cross sections are computed with a $10^{\circ}$ cut in the forward and backward scattering regions.  Similarly, a $1^{\circ}$ cut is applied for partial amplitudes, hence for $s^U$ determination. 

\begin{figure} 
\begin{center}	\includegraphics[width=0.95\textwidth]{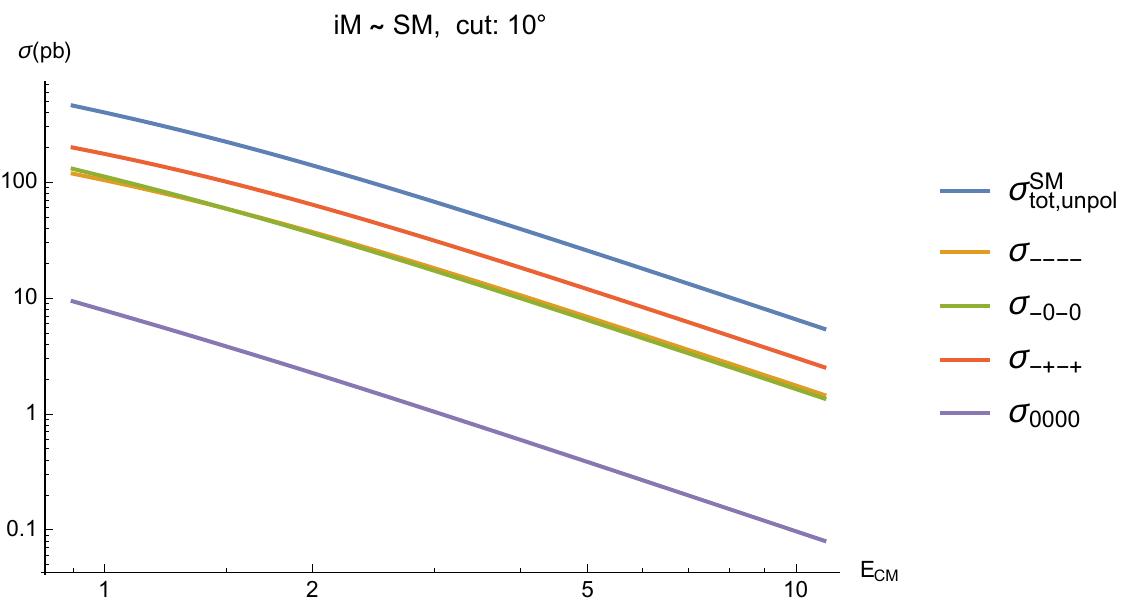}
\caption{Illustration of the contributions of different helicities (multiplicity taken into account) to the total unpolarized cross section as a function of the center-of-mass collision energy ($E_{CM} \equiv \sqrt{s}$, in TeV) in the SM.  The total cross section is shown in blue.}
\label{fig:plotlistPlot10DegreeParticPolsSMOnly}
\end{center}
\end{figure}

We begin by choosing  a set of independent helicity amplitudes for $W^+W^+$ scattering.  Altogether, there are 81 helicity amplitudes for this process but $P$ and $T$  discrete symmetries and the fact that the state has a symmetric wave function (Bose statistics) impose many relations between them and leave only 13 amplitudes as an independent set.  We choose them as follows:
\begin{equation}
\begin{array}{ccccccc}
	---- & ---0 & ---+ & --00 & --0+ & --++ & -0-0 \\ 
	-0-+ & -000 & -00+ & -+-+ & -+00 & 0000
\end{array}
\label{amplitudes}
\end{equation}
Here $+$, $-$ and $0$ denote the right-handed, left-handed and longitudinal
polarizations, respectively; the first two symbols define the initial state
and the last two symbols define the final state.
These amplitudes contribute to the total cross section with multiplicities 

\begin{displaymath}
\begin{array}{ccccccc}
	2 & 8 & 8 & 4 & 8 & 2 & 8 \\ 
	16 & 8 & 8 & 4 & 4 & 1 & 
\end{array}
\end{displaymath}
respectively, due to symmetry relations between all the 81 amplitudes.
In the SM their energy dependence is at most flat but their magnitude can differ by orders of magnitude (see Table \ref{fig:tabDraft1TeV} for their energy dependence and the contribution of the corresponding polarized cross sections to the total unpolarized cross section at 1 TeV).  The unpolarized cross section (decreasing like $1/s$) is saturated by just four of them, taking into account the corresponding multiplicities (see Fig.~\ref{fig:plotlistPlot10DegreeParticPolsSMOnly}).

The next thing of interest for us is the scattering energy $\sqrt{s^U}$ at which partial wave unitarity is violated by different helicity amplitudes for the two operators considered in this section, according to the tree level criterion $\left|\mathrm{Re}\{a_{J=0}\}\right|<1/2$.  This is shown in Table \ref{fig:tabEcmMax0Draft3} for the ${\cal O}_{S0}$ operator (positive $f$) and in Table \ref{fig:tabEcmMax0Draft3Ct0} for ${\cal O}_{T1}$ (negative $f$), as a function of the values of $f$.  We see that partial wave unitarity is first violated in the $0000$ amplitude for the first operator and in $----$ for the second one.  Unitarity is violated at vastly different energies for different helicity amplitudes, depending on the operator considered.  Some of them remain constant with energy, in particular some of those that saturate the SM total cross section.  The leading energy dependences of the amplitudes and the contributions of polarized cross sections to the total unpolarized cross section at the lowest $s^U$ where the first helicity amplitude violates partial wave unitarity are shown in Tables \ref{fig:tabDraftSUTrad} and \ref{fig:tabDraftSUTradCt0}, respectively.  One sees that for ${\cal O}_{S0}$, the $0000$ cross section (related to the amplitude which violates unitarity first) gives about 65\% of the total cross sections, independently of the value of $f$, for the corresponding values of minimal $s^U$.  For ${\cal O}_{T1}$, it is the $----$ cross section, closely followed by $--++$, with an about 80\% combined contribution to the total unpolarized cross section, independently of the value of $f$, for the corresponding values of minimal $s^U$.  The rest of the unpolarized cross sections at the minimal $s^U$ come (for both operators) from the helicity amplitudes that saturate the cross section in the SM, which either remain constant with energy (although weakly dependent on the value of $f$) or violate perturbative partial 
unitarity at a higher energy.

Unitarity bounds calculated from T-matrix diagonalization are virtually identical to
those for the amplitude that determines the minimal $s^U$ for ${\cal O}_{S0}$, while 
for ${\cal O}_{T1}$ they are about 15\% lower.

\begin{figure}[h]
\begin{center}	\includegraphics[width=0.9\textwidth]{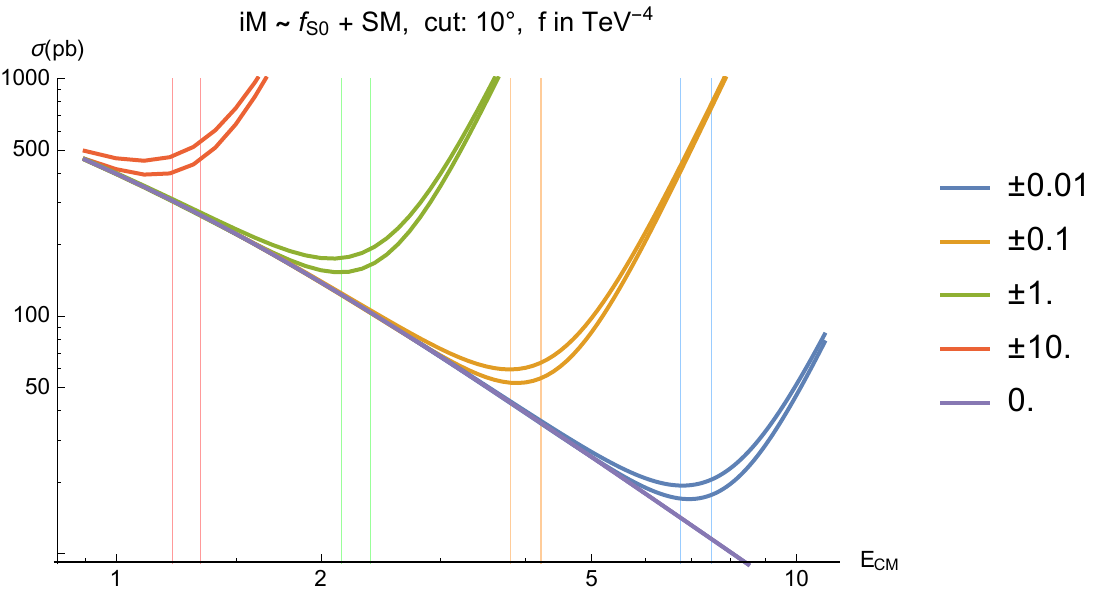}
\caption{Energy dependence of the total unpolarized $W^+W^+$ cross sections ($E_{CM} \equiv \sqrt{s}$, in TeV) for a chosen set of $f_{S0}$ values.  Vertical lines denote the lowest $\sqrt{s^U}$ for each value of $f$ (color correspondence).  There is no color distinction between the signs, upper lines (and hence stronger $\sqrt{s^U}$ limits) correspond to negative values of $f$.}
\label{fig:plotSigmaTotCs1GeqLeqGridZoom}
\end{center}
\end{figure}

\begin{figure}[h]
\begin{center}	\includegraphics[width=0.9\textwidth]{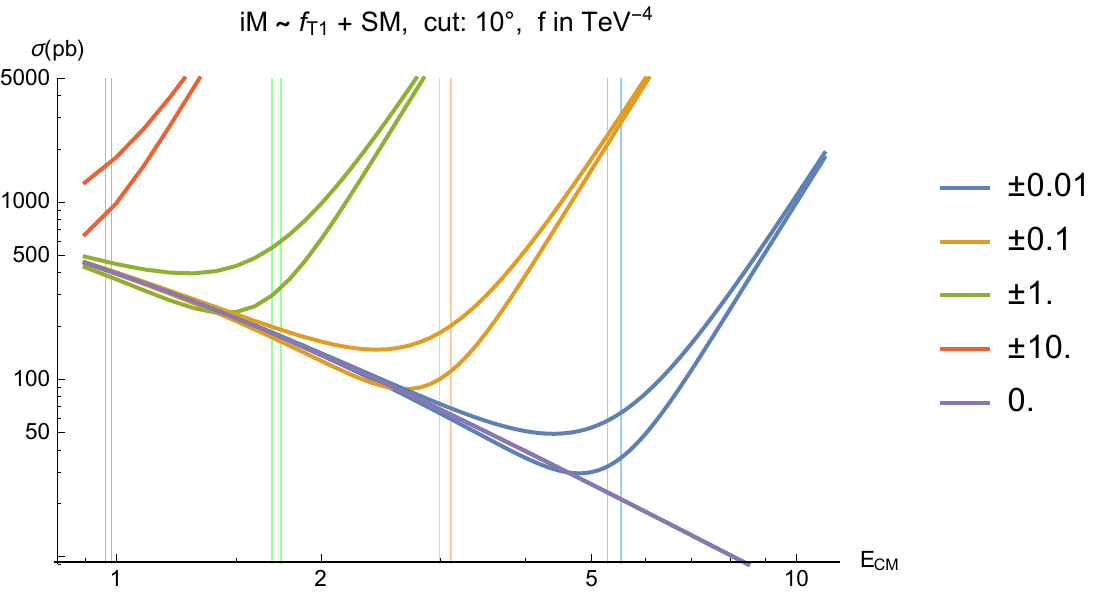}
\caption{Energy dependence of the total unpolarized $W^+W^+$ cross sections ($E_{CM} \equiv \sqrt{s}$, in TeV) for a chosen set of $f_{T1}$ values.  Vertical lines denote the lowest $\sqrt{s^U}$ for each value of $f$ (color correspondence).  There is no color distinction between the signs, upper lines (and hence stronger $\sqrt{s^U}$ limits) correspond to negative values of $f$.}
\label{fig:plotSigmaTotCt0GeqLeqGridZoom}
\end{center}
\end{figure}

\begin{figure}[h]
\begin{center}	\includegraphics[width=0.95\textwidth]{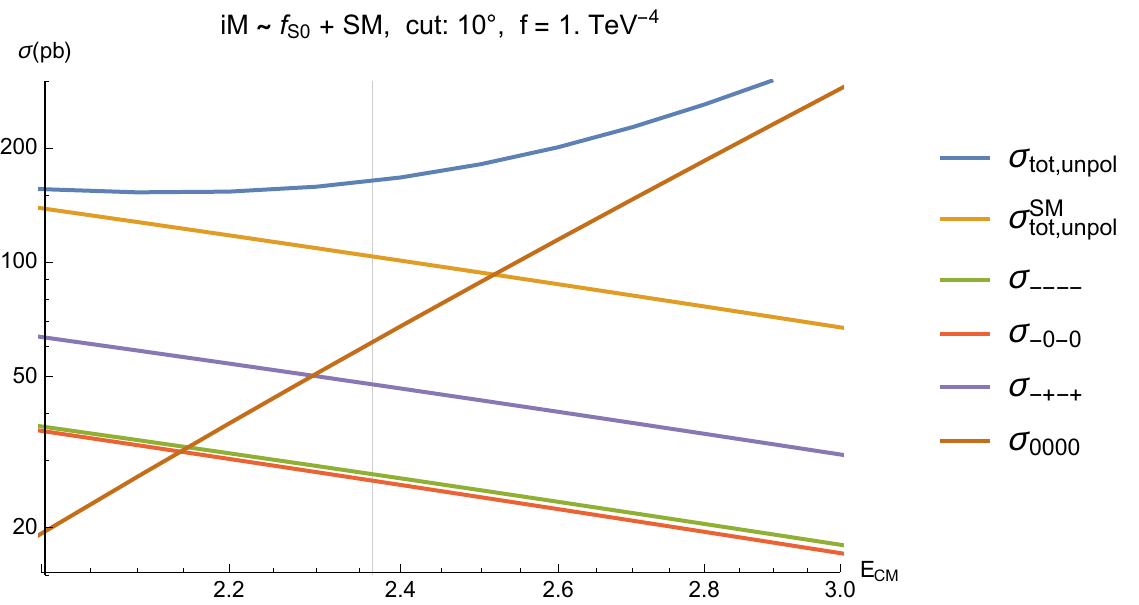}
\caption{Contributions of the polarized cross sections (multiplicity taken into account) to the total unpolarized cross section as a function of the center-of-mass collision energy
($E_{CM} \equiv \sqrt{s}$, in TeV) for $f_{S0} = 1.  \ \mathrm{TeV}^{-4}$.  The total cross section is shown in blue, the total cross section in the SM is shown in orange.  The remaining polarized cross sections are negligibly small.}

\label{fig:plotlistPlot10DegreeParticPolsfEq1}
\end{center}
\end{figure}

\begin{figure}[h]
\begin{center}	\includegraphics[width=0.95\textwidth]{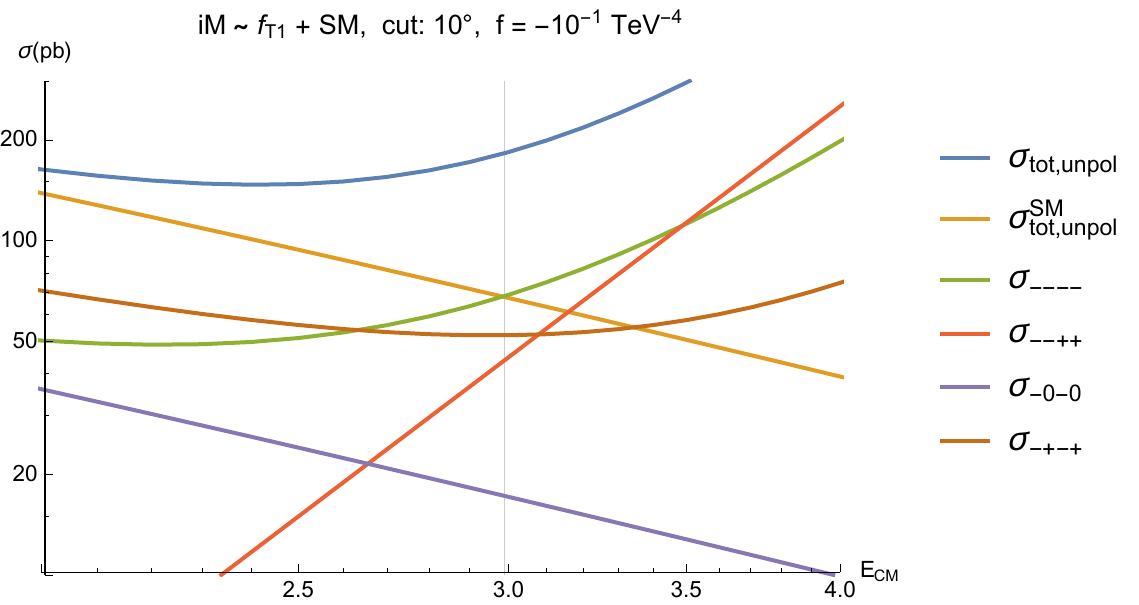}
\caption{Contributions of the polarized cross sections (multiplicity taken into account) to the total unpolarized cross section as a function of the center-of-mass collision energy
($E_{CM} \equiv \sqrt{s}$, in TeV) for $f_{T1} = -0.1 \ \mathrm{TeV}^{-4}$.  The total cross section in shown in blue, the total cross section in the SM is shown in orange.  The remaining polarized cross sections are negligibly small.}
\label{fig:plotlistPlot10DegreeParticPolsft0Eq1Zoom}
\end{center}
\end{figure}

Some examples of the energy dependence of the cross sections for both operators are shown in the following figures: for the total unpolarized cross sections with ${\cal O}_{S0}$ in  Fig.~\ref{fig:plotSigmaTotCs1GeqLeqGridZoom}, for polarized cross sections with ${\cal O}_{S0}$ in Fig.~\ref{fig:plotlistPlot10DegreeParticPolsfEq1}, for unpolarized with ${\cal O}_{T1}$ in Fig.~\ref{fig:plotSigmaTotCt0GeqLeqGridZoom}, and for polarized with ${\cal O}_{T1}$ in Fig.~\ref{fig:plotlistPlot10DegreeParticPolsft0Eq1Zoom}.  We observe that both operators show several similar interesting features.  Below the partial wave unitarity minimal bounds $s^U$, sizable deviations from the SM predictions occur only for small energy intervals close to those bounds.  
This is the region where the quadratic term in Eq.~\eqref{full1} begins to dominate BSM effects
(see Figs.~\ref{fig:plotSigmaTotCs1GeqLeqGridZoom} and \ref{fig:plotSigmaTotCt0GeqLeqGridZoom}).
The contribution of the interference term at this point generally depends on
which helicity combinations get affected by a given operator and how much they
contribute to the total cross section in the SM.  Interference is visible
for ${\cal O}_{S0}$, ${\cal O}_{S1}$, ${\cal O}_{T0}$, ${\cal O}_{T1}$, ${\cal O}_{T2}$,
${\cal O}_{M1}$, ${\cal O}_{M7}$, and negligible for ${\cal O}_{M0}$ and ${\cal O}_{M6}$.
The energy dependence of the unpolarized cross sections around the values of the minimal $s^U$ is moreover somewhat weakened by the contribution from the  helicity amplitudes that have not reached the unitarity limit.

\vspace{2cm}

\begin{table}[h]
\begin{center}	\includegraphics[width=0.95\textwidth]{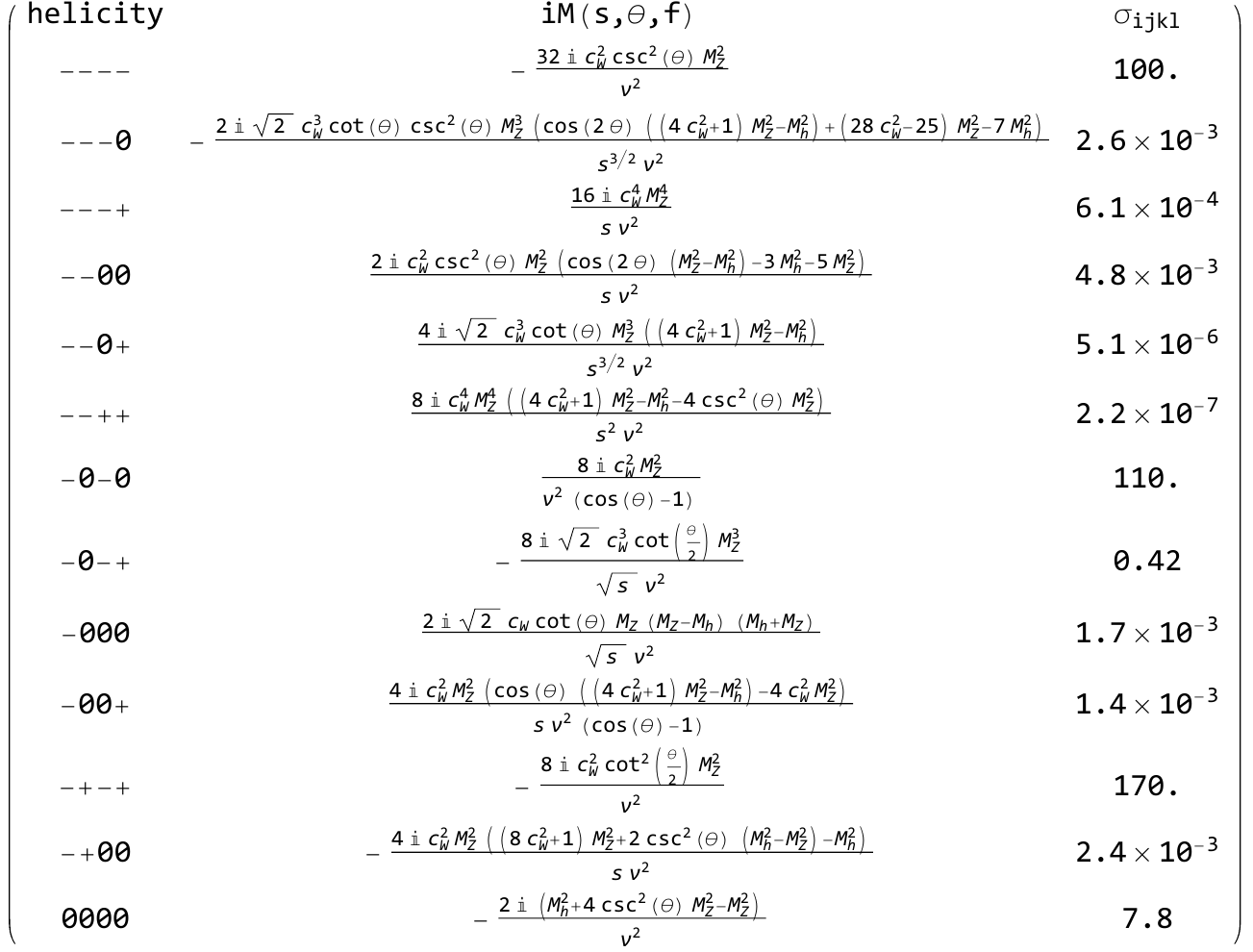}
\caption{The leading energy behavior in the limit $s>>(\mathrm{EW\  scale})^2$ of the scattering amplitude $iM$ for all the 13 independent helicities in the SM case.  In the third column shown are numerical values (in pb) of the contributions from the helicities to the total unpolarized cross section at 1 TeV; $c_W$ is the cosine of the Weinberg mixing angle, $v$ is the SM Higgs vev, $\theta$ is the scattering angle.}
\label{fig:tabDraft1TeV}
\end{center}
\end{table}

\begin{table} 
\begin{center}	\includegraphics[width=.55\textwidth]{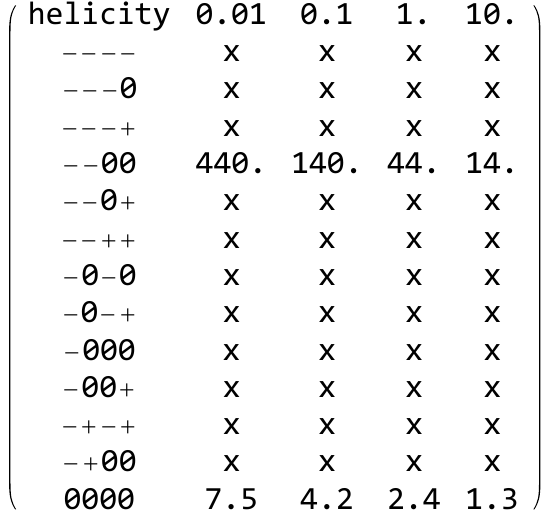}
\caption{Values of $\sqrt{s^U}$ (in TeV) for all the helicity amplitudes from the partial wave unitarity criterion for a chosen set of $f_{S0}$ values (first row, in $\mathrm{TeV}^{-4}$); ``x'' denotes no unitarity violation.}
%the cases where there is no growth of the amplitude with energy and thus the unitarity condition is identically satisfied.}
\label{fig:tabEcmMax0Draft3}
\end{center}
\end{table}

\begin{table} 
\begin{center}	\includegraphics[width=.8\textwidth]{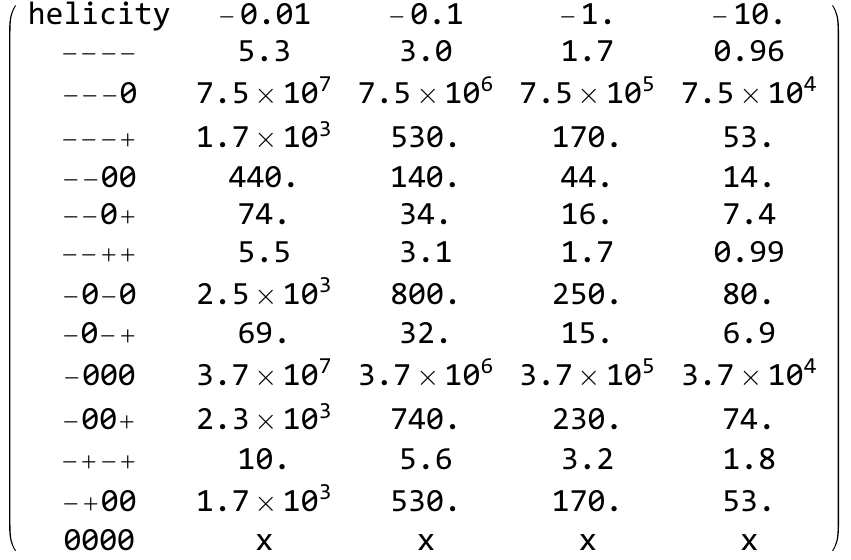}
\caption{Values of $\sqrt{s^U}$ (in TeV) for all the helicity amplitudes from the partial wave unitarity criterion for a chosen set of $f_{T1}$ values (first row, in $\mathrm{TeV}^{-4}$); ``x'' denotes no unitarity violation.}
%the cases where there is no growth of the amplitude with energy and thus the unitarity condition is identically satisfied.}
\label{fig:tabEcmMax0Draft3Ct0}
\end{center}
\end{table}

\begin{table} 
\begin{center}	\includegraphics[width=1.1\textwidth]{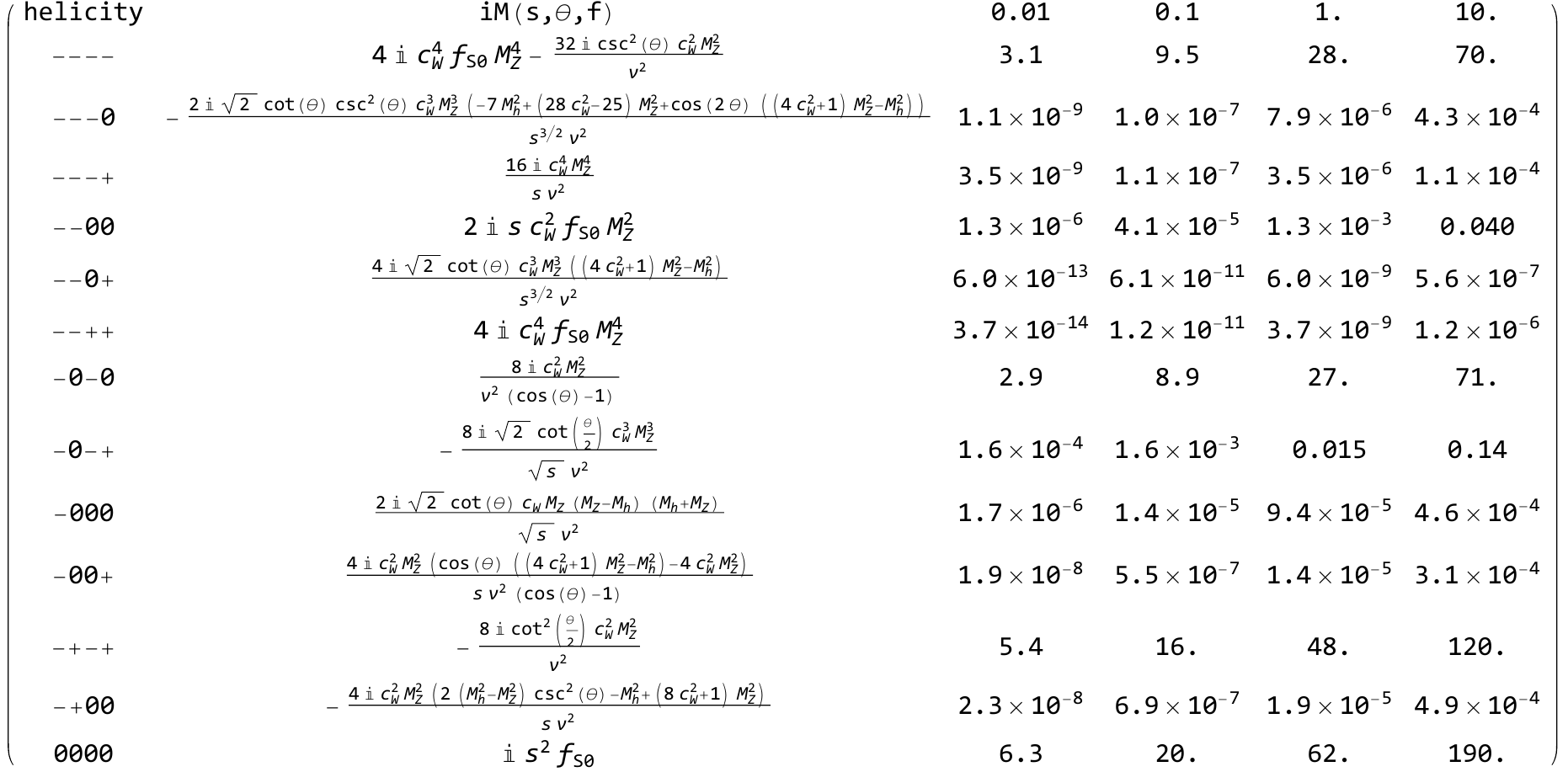}
\caption{The leading energy dependence of the amplitudes (conventions as in Fig.~\ref{fig:tabDraft1TeV}) and the contribution of the polarized cross sections (in pb) to the total unpolarized cross sections at the minimal $s^U$ for a chosen set of $f_{S0}$ values (first row, in $\mathrm{TeV}^{-4}$).}
\label{fig:tabDraftSUTrad}
\end{center}
\end{table}

\begin{table} 
\begin{center}	\includegraphics[width=1.1\textwidth]{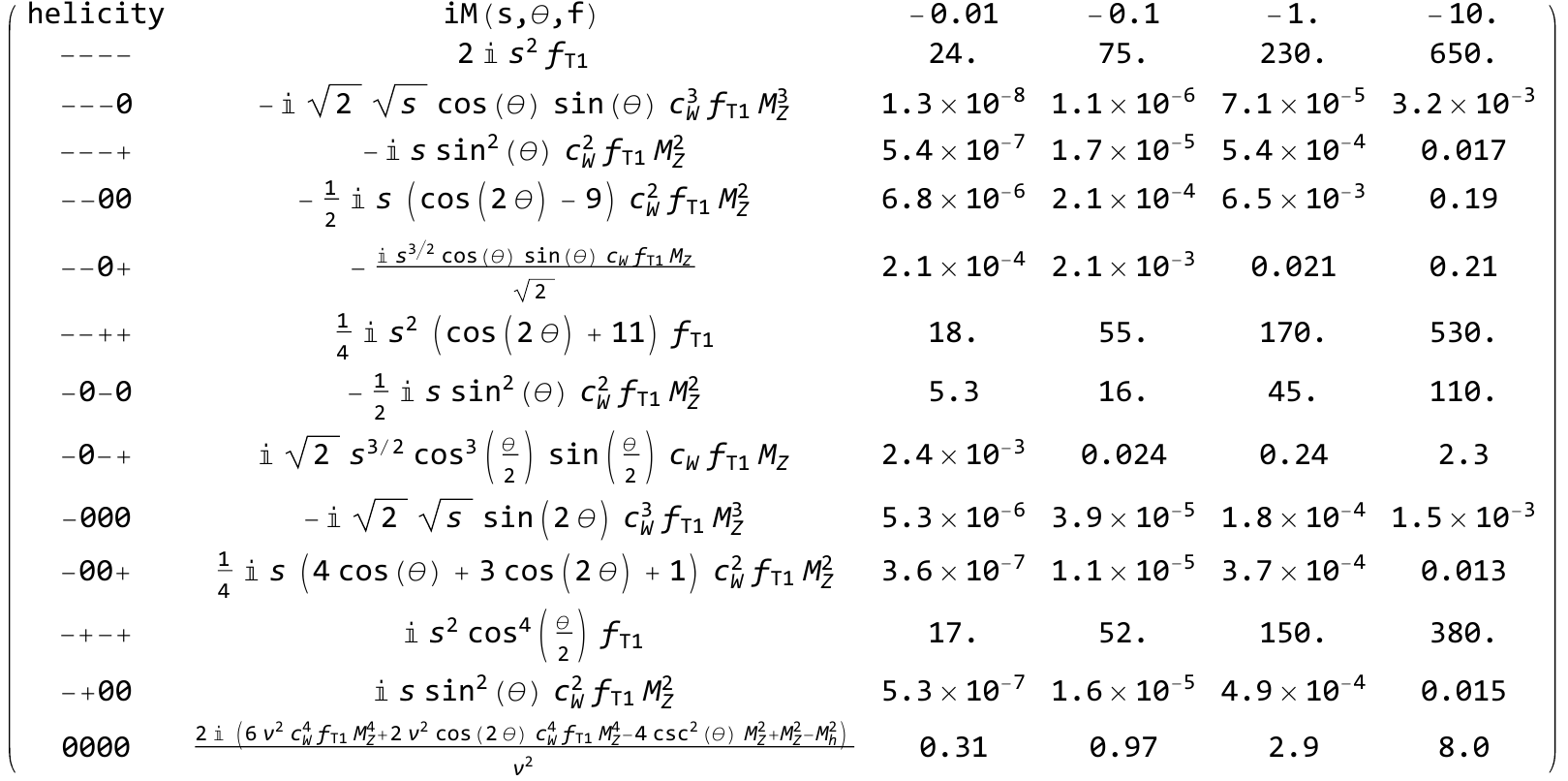}
\caption{The leading energy dependence of the amplitudes (conventions as in Fig.~\ref{fig:tabDraft1TeV}) and the contribution of the polarized cross sections (in pb) to the total unpolarized cross sections at the minimal $s^U$ for a chosen set of $f_{T1}$ values (first row, in $\mathrm{TeV}^{-4}$). }
\label{fig:tabDraftSUTradCt0}
\end{center}
\end{table}

%\begin{figure} 
%\begin{center}	\includegraphics[width=0.5\textwidth]{nonDiagVsDiagMat2.pdf}
%\caption{''diag.'' oznacza $\sqrt{s^U}$ z diagonalizacji; ''no diag.'' z najsilniejszej skretnosci; $f$ w $\mathrm{TeV}^{-4}$, $\sqrt{s^U}$ w TeV}
%\label{fig:nonDiagVsDiagMat2}
%\end{center}
%\end{figure}

%\end{multicols}a

\clearpage

%%%%%%%%%%%%%%%%%%%%%%%%%%%%%%%%%%%%%%%%%%%%%%%%%%%%%%%%%%%%%%%%%%%%%%%%%%%%%%%%%%%%%%%%%%%%%%%%%%%%%%%%%%
% BIBLIOGRAPHY
% REMEMBER TO WRITE THE LINE bibtex article IN THE TERMINAL TO GENERATE THE CORRESPONDING FILES
%\bibliography{}
%%%%%%%%%%%%%%%%%%%%%%%%%%%%

\end{document}